\documentclass[fleqn,usenatbib]{mnras}

\usepackage{newtxtext,newtxmath}
\usepackage[T1]{fontenc}
\usepackage{ae,aecompl}
\usepackage{graphicx}
\usepackage{amsmath,braket,bm,aas_macros,ascmac}
\usepackage{fancybox}
\numberwithin{equation}{section}

\title[Relativistic distortions in GI correlations]%
{Relativistic distortions in galaxy density-ellipticity correlations: gravitational redshift and peculiar velocity effects
}
\author[S. Saga et al.]
{Shohei Saga,$^{1,2}$\thanks{E-mail: saga@iap.fr}
Teppei Okumura,$^{3,4}$
Atsushi Taruya$^{5,4}$ and
Takuya Inoue$^{3,6}$
\\
$^{1}$Sorbonne Universti\'e, CNRS, UMR7095, Institut d'Astrophysique de Paris, 98bis boulevard Arago, F-75014 Paris, France \\
$^{2}$Laboratoire Univers et Th{\'e}ories, Observatoire de Paris, Universit{\'e} PSL, CNRS, Universit{\'e} de Paris, 5 place Jules Janssen 92190 Meudon, France \\
$^{3}$Academia Sinica Institute of Astronomy and Astrophysics (ASIAA), No. 1, Section 4, Roosevelt Road, Taipei 10617, Taiwan \\
$^{4}$Kavli Institute for the Physics and Mathematics of the Universe (WPI), The University of Tokyo Institutes for Advanced Study, \\
The University of Tokyo, 5-1-5 Kashiwanoha, Kashiwa, Chiba 277-8583, Japan \\
$^{5}$Center for Gravitational Physics and Quantum Information, Yukawa Institute for Theoretical Physics, Kyoto University, Kyoto 606-8502, Japan\\
$^{6}$Department of Physics, National Taiwan University, No. 1, Section 4, Roosevelt Road, Taipei 10617, Taiwan\\
}
\thisfancyput(14.5cm,0.5cm){\large{YITP-22-69}}
\date{Accepted XXX. Received YYY; in original form ZZZ}
\pubyear{2022}
\begin{document}
\label{firstpage}
\pagerange{\pageref{firstpage}--\pageref{lastpage}}
\maketitle
\begin{abstract}
We study relativistic effects, arising from the light propagation in an inhomogeneous universe. We particularly investigate the effects imprinted in a cross-correlation function between galaxy positions and intrinsic galaxy shapes (GI correlation).
Considering the Doppler and gravitational redshift effects as major relativistic effects, we present an analytical model of the GI correlation function, from which we find that the relativistic effects induce non-vanishing odd multipole anisotropies.
Focusing particularly on the dipole anisotropy, we show that the Doppler effect dominates at large scales, while the gravitational redshift effect originated from the halo potential dominates at the scales below $10$--$30\, {\rm Mpc}/h$, with the amplitude of the dipole GI correlation being positive over all the scales.
Also, we newly derive the covariance matrix for the modelled GI dipole.
Taking into account the full covariance, we estimate the signal-to-noise ratio and show that the GI dipole induced by the relativistic effects is detectable in future large-volume galaxy surveys.
We discuss how the measurement of dipole GI correlation could be helpful to detect relativistic effects in combination with the conventional galaxy-galaxy cross correlation.
\end{abstract}
\begin{keywords}
Cosmology -- large-scale structure of Universe --  dark matter
\end{keywords}

\section{Introduction}

The spatial distribution of galaxies, which contains a wealth of cosmological information, has been extensively studied via galaxy redshift surveys for decades. Via the spectroscopic measurements, the observed density field appears distorted mainly due to the Doppler effect induced by the peculiar velocities of galaxies. It results in apparent anisotropies along the line-of-sight direction, known as redshift-space distortions \citep[RSD; ][]{1987MNRAS.227....1K,1992ApJ...385L...5H}.
The RSD of galaxy clustering is now recognised as a sensitive probe of the growth of the structure, and hence future observations are intently planned.

There are other special and general relativistic effects on observed galaxy clustering on top of the Doppler effects, e.g., gravitational redshift, integrated Sachs-Wolfe, and weak lensing effects~\citep[e.g.,][]{1987MNRAS.228..653S,2000ApJ...537L..77M,2004MNRAS.348..581P,2009PhRvD..80h3514Y,2010PhRvD..82h3508Y,2011PhRvD..84f3505B,2011PhRvD..84d3516C,2012JCAP...10..025B,2014CQGra..31w4001Y}.
These effects produce asymmetric distortions to the galaxy distribution along the line-of-sight direction~\citep{2013MNRAS.434.3008C,2012arXiv1206.5809Y,2018JCAP...03..019T}.
Applying the multipole expansion with a certain definition of the line-of-sight direction, the asymmetry in the galaxy-galaxy correlation function (GG correlation) is characterized by the non-vanishing odd multipole moments, among which the dipole moment gives the largest amplitude~\citep[e.g.,][for the linear regime]{2009JCAP...11..026M,2014PhRvD..89h3535B}.
Beyond the linear regime, the behaviour of the dipole moment has been studied both analytically~\citep{2019JCAP...04..050D,2020JCAP...07..048B,2020MNRAS.498..981S} and numerically~\citep{2019MNRAS.483.2671B,2017MNRAS.471.3899B,2021MNRAS.501.2547G,2021MNRAS.504.3534C}.
According to recent findings~\citep{2019MNRAS.483.2671B,2020MNRAS.498..981S}, the GG dipole is dominated at large scales by the Doppler effect through the corrections beyond the plane-parallel limit. On the other hand, at small scales ($\lesssim30-40$Mpc$/h$), the dipole signal is dominated by the gravitational redshift effect induced by the halo potential, with the sign of the dipole amplitude flipped.
Thus, measuring the GG dipole particularly at small scales offers a unique opportunity to test the theory of gravity through the gravitational redshift effects \citep[see e.g.,][for the test of gravity from the viewpoint of equivalence principle]{2018JCAP...05..061B,2020JCAP...08..004B}.

In the GG correlations, cross-correlating two different biased objects is essential in detecting the non-vanishing dipole moment\footnote{It is not the case when a higher-order correlation is considered. See \cite{2019MNRAS.486L.101C} and \cite{2020JCAP...03..065M} for the detectability of relativistic effects using the galaxy bispectrum with a single galaxy population.}.
Thus, given one galaxy sample in a galaxy survey, we need to split the sample into at least two subsamples. Depending on how we split the sample, the detectability drastically changes~\citep{2022MNRAS.511.2732S}, and we need to find an optimal way to measure the GG dipole at a statistically significant level~\citep[see e.g.,][]{2016JCAP...08..021B,2018JCAP...05..043L}.

To avoid this, one can consider to use additional information on the intrinsic property of galaxies, whose statistical nature is related to the large-scale matter distribution, and to cross-correlate it with galaxy density fields.
One representative example of such an intrinsic property is the galaxy shape. In weak gravitational lensing analysis, a coherent alignment of galaxy shapes, i.e., intrinsic alignments (IAs), has been known as a systematic contaminant ~(see e.g., \citealt{2000MNRAS.319..649H,2000ApJ...545..561C,2001MNRAS.323..713C,2001ApJ...559..552C,2004PhRvD..70f3526H,2006MNRAS.367..611M,2007MNRAS.381.1197H,2009ApJ...694..214O}, and a review \citealt{2015PhR...558....1T}). The correlation function between galaxy positions and intrinsic shapes (GI correlations) has been clearly detected in the observation in both two-dimensional projected sky and three-dimensional redshift space~(e.g., \citealt{2006MNRAS.367..611M,2007MNRAS.381.1197H,2009ApJ...694L..83O,2015MNRAS.450.2195S,2011A&A...527A..26J,2019A&A...624A..30J,2019MNRAS.489.5453S,2020ApJ...904..135Y}), even at higher redshifts of $z>1$ \citep{2022ApJ...924L...3T}. 
Nevertheless, IAs have been recently recognised as a new probe of cosmological models~\citep{2012PhRvD..86h3513S,2013JCAP...12..029C,2019PhRvD.100j3507O,2020MNRAS.493L.124O,2020ApJ...891L..42T,2021MNRAS.503L...6S}.
Besides, combining the conventional GG correlation with IA correlations is beneficial, leading to a tight constraint on the cosmological parameters \citep{2020ApJ...891L..42T,2022PhRvD.106d3523O,2022MNRAS.515.4464C}.

This paper aims to study a possible imprint of the gravitational redshift on the GI correlation and to discuss its detectability.
In doing so, we present an analytical model of the GI correlation including two major relativistic effects, namely the gravitational redshift and Doppler effects, taking also the wide-angle effect into consideration.
The model presented here combines the quasi-linear treatment of the observed density field, which has been exploited in our previous work \citep{2020MNRAS.498..981S,2022MNRAS.511.2732S}, with the linear alignment model for the galaxy intrinsic shape \citep{2001MNRAS.320L...7C,2004PhRvD..70f3526H}.
To quantitatively estimate the signal-to-noise ratio of the GI dipole, 
we also derive the analytical expressions for the covariance error matrix that properly accounts for the angular dependence of the GI correlations.
We then compute the signal-to-noise ratio of the GI dipole and investigate its dependence on several key parameters.

This paper is organized as follows.
In Sec.~\ref{sec: preliminary}, we introduce the model of the observed density fields including relativistic corrections and ellipticity fields. With these models, Sec.~\ref{sec: GI corr} presents the analytical model of the GI correlation, taking the Doppler and gravitational redshift effects into account.
In Sec.~\ref{sec: detectability}, for a quantitative estimate of the signal-to-noise ratio, we present the covariance matrix for the GI dipole correlation function, leaving a detailed derivation to Appendix~\ref{app: covariance}. We then compute the signal-to-noise ratio for a representative setup of the upcoming surveys. We show that in contrast to the GG dipole, contributions to the signal mainly come from large scales, where the Doppler effect dominates the GI dipole, with the signal-to-noise ratio typically (S/N)$\, \{1\,({\rm Gpc}/h)^3/V\}^{1/2}\sim\mathcal{O}(1)$. However, we find that the signal-to-noise ratio is further enhanced at small scales by the gravitational redshift effect from the halo potential, thus improving the detectability of the GI dipole. Finally, Sec.~\ref{sec: summary} is devoted to the summary of important findings.
In Appendix~\ref{app: coefficients}, we summarize the relevant coefficients involved in the expressions of our analytical model.

Throughout this paper, we assume a flat Lambda cold dark matter model. We choose the fiducial values of cosmological parameters to match the seven-year WMAP results~\citep{2011ApJS..192...18K}, and we will work with units of $c=1$.

\section{Preliminaries}
\label{sec: preliminary}

In this paper, as a new tool to study relativistic effects, we focus on the three-dimensional GI correlation function. It is defined by the cross-correlation between the density ($\delta$) and projected ellipticity fields ($\gamma_{+/\times}$) of galaxies, respectively observed at positions $\bm{s}_{1}$ and $\bm{s}_{2}$ in redshift space:
\begin{align}
\xi^{\delta\gamma_{+/\times}}(\bm{s}_{1}, \bm{s}_{2}) &\equiv \Braket{\delta(\bm{s}_{1})\gamma_{+/\times}(\bm{s}_{2})} \label{eq: def GI corr} ~,
\end{align}
where the bracket $\Braket{\cdots}$ stands for the ensemble average.
In this section, we present the model of the two cosmological fields ($\delta$ and $\gamma$) in the GI correlation function.
We first write down the relation between redshift-space and real-space positions of galaxies in Sec.~\ref{sec:gr} under the presence of relativistic effects.
In Sec.~\ref{sec: delta RSD}, we briefly describe the analytical model for the observed galaxy density field, $\delta(\bm{s})$, which takes into account the major relativistic contributions based on \citet{2022MNRAS.511.2732S}.
We then consider the ellipticity of galaxies, $\gamma$, as a tracer of the underlying gravitational tidal field. Based on the linear alignment model, we give an explicit relation between the ellipticity field and surrounding matter distribution in the large-scale structure in Sec.~\ref{sec: LA model}.

\subsection{Relativistic effects}
\label{sec:gr}

The spatial distribution of galaxies observed in redshift surveys is distorted due to the special and general relativistic effects \citep[e.g.,][and references therein]{2013MNRAS.434.3008C,2014CQGra..31w4001Y,2018JCAP...03..019T}.
A well-known source of such distortions is the Doppler effect due to the peculiar velocity of galaxies \citep{1987MNRAS.227....1K,1992ApJ...385L...5H}. 
On top of this, galaxy clustering appears asymmetric in the presence of other relativistic effects, e.g., the gravitational redshift, integrated Sachs-Wolfe, Shapiro time delay, transverse Doppler, and weak lensing effects~\citep{2009JCAP...11..026M,2011PhRvD..84f3505B,2014PhRvD..89h3535B}. Among them, we have recently shown that the gravitational redshift effect is the dominant source of the dipole anisotropy at small scales~\citep{2019MNRAS.483.2671B,2020MNRAS.498..981S}.
Thus, in this paper, we focus on the Doppler and gravitational redshift effects, ignoring other subdominant contributions.

By solving the geodesic equation in the presence of matter inhomogeneities, 
the observed source position distorted by the Doppler and gravitational redshift effects in redshift space ($\bm{s}$) is related to the real-space counterpart ($\bm{x}$) as
\begin{align}
\bm{s} &= \bm{x} + \frac{1}{aH}\left( \bm{v}\cdot \hat{\bm{x}}\right)\hat{\bm{x}}
-\frac{1}{aH}
\phi(\bm{x})
\hat{\bm{x}} ~,
\label{eq: s to x}
\end{align}
where the hat denotes a unit vector and the quantities $a$, $H$, $\bm{v}$, and $\phi(\bm{x})$ are a scale factor, Hubble parameter, peculiar velocity of galaxies, and the gravitational potential, respectively.
The second term of the right-hand side represents the Doppler effect including the wide-angle effect~(e.g.,~\citealt{1996MNRAS.278...73H,1998ApJ...498L...1S,2000ApJ...535....1M,2004ApJ...614...51S}) and the third term represents the gravitational redshift effect caused by the gravitational potential.
This equation is valid in the weak-field approximation of the metric perturbations and $|\bm{v}| \ll 1$.
The explicit form of other minor contributions are found in e.g.,~\citet{2010PhRvD..82h3508Y,2011PhRvD..84d3516C,2011PhRvD..84f3505B}.

Photons emitted from galaxies generally come from the bottom of deep gravitational potentials of the host dark matter haloes. The gravitational redshift due to the halo potential thus contributes significantly to the observed redshift.
Since haloes are formed through non-linear physical processes, the halo potential at the galaxy position cannot be simply characterized by the linear density field. 
Because of this fact, \citet{2020MNRAS.498..981S,2022MNRAS.511.2732S} modelled the gravitational redshift effect by splitting it into contributions from the linear gravitational potential, $\phi_{\rm L}$, and the non-perturbative halo potential, $\phi_{\rm halo}$\footnote{The halo gravitational potential, $\phi_{\rm halo}$, is related to the quantity $\epsilon_{\rm NL}$ defined in \citet{2022MNRAS.511.2732S}
as $\epsilon_{\rm NL} = -\phi_{\rm halo}/(aH)$.}:
\begin{align}
\phi(\bm{x}) = \phi_{\rm L}(\bm{x}) + \phi_{\rm halo} .
\end{align}
Following \citet{2020MNRAS.498..981S,2022MNRAS.511.2732S}, we adopt the Navarro-Frenk-White~(NFW) density profile~\citep{1996ApJ...462..563N} to describe this non-perturbative contribution: $\phi_{\rm halo} = \phi_{\rm NFW,0}(z,M)$, with the function $\phi_{\rm NFW,0}(z,M)$ being the NFW potential evaluated at the centre of haloes~\citep[see Appendix~D of][for the explicit form]{2020MNRAS.498..981S}.
In general, the most massive or brightest galaxy in a halo tends to reside in the centre of the halo. The position is, however, not necessarily the bottom of the gravitational potential and sometimes off-centred~\citep{2013MNRAS.435.2345H,2020MNRAS.493.1120Y}.
This off-centering effect affects the estimation of the non-perturbative term.
Since this paper mainly aims to investigate the basic features of the relativistic effects on the GI correlation, we do not take into account this systematic effect and leave a detailed investigation for future work.

\subsection{Galaxy density field}
\label{sec: delta RSD}

Based on the mapping relation at Eq.~(\ref{eq: s to x}), \citet{2020MNRAS.498..981S,2022MNRAS.511.2732S} have derived the analytical expression for the redshift-space galaxy density field including both the Doppler and gravitational redshift effects. Shortly, we employed the first-order Lagrangian perturbation theory known as the Zel'dovich approximation~\citep{1969JETP...30..512N,1970A&A.....5...84Z,1989RvMP...61..185S}, and obtained the expression for the density field by keeping and expanding the terms up to $O(\phi_{\rm halo} \delta_{\rm L})$. The result is summarized as follows:
\begin{align}
\delta(\bm{s}) &=
\delta^{(\rm r)} (\bm{s}) + \delta^{(\rm Dop)} (\bm{s}) + \delta^{(\rm grav)} (\bm{s}) + \delta^{(\rm nl)}(\bm{s}) ~. \label{eq: delta std + delta epsNL}
\end{align}
Here, the galaxy density field in redshift space is decomposed into four contributions: the real-space density field, $\delta^{(\rm r)} (\bm{s})$, the Doppler effect, $\delta^{(\rm Dop)}(\bm{s})$, 
and the gravitational redshift effects due to the linear and non-perturbative halo potentials, $\delta^{(\rm grav)}(\bm{s})$ and $\delta^{(\rm nl)}(\bm{s})$, respectively.
These contributions are all expressed in terms of the linear galaxy density field, explicitly given by
\begin{align}
\delta^{(\rm r)}(\bm{s}) &=  \int\frac{{\rm d}^{3}\bm{k}}{(2\pi)^{3}}{\rm e}^{{\rm i}\bm{k}\cdot\bm{s}}
b \delta_{\rm L}(\bm{k})
\label{eq: def delta real}
~, \\
\delta^{(\rm Dop)}(\bm{s}) &=  \int\frac{{\rm d}^{3}\bm{k}}{(2\pi)^{3}}{\rm e}^{{\rm i}\bm{k}\cdot\bm{s}}
\Biggl[ f\mu_{k}^{2} - {\rm i} f \frac{2}{ks}\mu_{k} \Biggr]\delta_{\rm L}(\bm{k})
\label{eq: def delta std}
~, \\
\delta^{(\rm grav)}(\bm{s}) &=  \int\frac{{\rm d}^{3}\bm{k}}{(2\pi)^{3}}{\rm e}^{{\rm i}\bm{k}\cdot\bm{s}}
\Biggl[ \left( iks\mu_{k} + 2\right) \frac{\mathcal{M}}{sk^{2}} \Biggr]\delta_{\rm L}(\bm{k})
\label{eq: def delta grav}
~, \\
\delta^{(\rm nl)}(\bm{s}) & = 
-\frac{\phi_{\rm halo}}{aHs}\int\frac{{\rm d}^{3}\bm{k}}{(2\pi)^{3}}{\rm e}^{{\rm i}\bm{k}\cdot\bm{s}}
\Biggl[
-1 + (1-2f)\mu^{2}_{k} 
\notag \\
& \quad
- {\rm i}(1+f)\frac{2}{ks}\mu_{k} - {\rm i}b  ks \mu_{k} - {\rm i} f ks \mu^{3}_{k}
\Biggr]\delta_{\rm L}(\bm{k}) ~,
\label{eq: def delta epsNL}
\end{align}
where the quantities $\delta_{\rm L}(\bm{k})$, $b$, and $f$ are, respectively, the Fourier counterpart of the linear galaxy density field, the Eulerian linear galaxy bias and the linear growth rate defined by $f \equiv {\rm d}\ln{D_{+}(a)}/{\rm d}\ln{a}$ with $D_{+}$ being the linear growth factor.
The quantity $\mu_k$ is the directional cosine between the line-of-sight and wave vectors, $\mu_{k}\equiv \hat{\bm{s}}\cdot\hat{\bm{k}}$. The time-dependent function $\mathcal{M}$ is defined by $\mathcal{M} \equiv - 3\Omega_{\rm m0}H^{2}_{0}/(2a^{2}H)$ with $\Omega_{m0}$ and $H_{0}$ being the matter density parameter and Hubble parameter at the present time, respectively.

Eq.~(\ref{eq: delta std + delta epsNL}) with Eqs.~(\ref{eq: def delta real})--(\ref{eq: def delta epsNL}) is the key expression for the model of the galaxy density field in redshift space, accounting for the gravitational redshift effect induced by the halo potential.
\citet{2020MNRAS.498..981S,2022MNRAS.511.2732S} have shown that this model well describes the dipole moment of halo-halo cross correlation measured in simulations down to $5\,{\rm Mpc}/h$.
In particular, the non-perturbative term in Eq.~(\ref{eq: def delta epsNL}) successfully explains the sign-flip behaviour of the GG dipole at small scales ($s\approx 30\, {\rm Mpc}/h$) seen in the simulations.
We will use this model to qualitatively investigate the behaviour of the GI correlation function, and its detectability.

Note that in deriving Eq.~(\ref{eq: delta std + delta epsNL}), we have ignored the magnification bias, i.e., the apparent density fluctuations through the fluctuation in luminosity distances caused by flux-limited galaxy samples.
While the magnification bias is also known to induce the asymmetric galaxy clustering beyond the plane-parallel limit~\citep[e.g.,][]{2011PhRvD..84f3505B,2017PhRvD..95d3530H}, its contribution to the GG dipole is sub-dominant compared to the non-perturbative contribution, especially at high redshift~\citep{2022MNRAS.511.2732S}.
We will address the magnification bias effect on the GI correlation in future investigations together with other potential systematic effects, e.g., the transverse Doppler effect.

\subsection{Galaxy intrinsic ellipticity field}
\label{sec: LA model}

In this paper, we use the ellipticity of galaxies, $\gamma_{+/\times}(\bm{s})$, as a tracer of the underlying gravitational tidal field arising from the surrounding matter distribution in the large-scale structure.
In general, the observed ellipticity of galaxies is divided into two contributions in the presence of the lensing effect, the intrinsic ellipticity and lens-induced ellipticity. Throughout the paper, we will consider the cross-correlation between density and ellipticity field in the same redshift slice. In this case, the contribution from the lens-induced ellipticity is small and can be ignored~\citep[e.g.,][]{2000ApJ...537L..77M,2008PhRvD..77f3526H,2022PhRvD.106d3523O}. Hence, we consider only the contribution of the intrinsic ellipticity.

To compute the GI correlation function, we need a model which relates the intrinsic ellipticity to the density field.
In this paper, we adopt the linear alignment model~\citep{2001MNRAS.320L...7C,2004PhRvD..70f3526H}, in which the intrinsic ellipticity is linearly related to the real-space density field.
We formulate the model without assuming the plane-parallel limit. We can therefore use the resultant expression for widely separated pairs~\citep[see][for the formulation based on the different basis]{2021MNRAS.503L...6S}.

Let us denote the traceless part of the second moment of the intrinsic galaxy shape, defined in three-dimensional space, by $\gamma^{\rm I}_{ij}(\bm{x})$. In the linear alignment model, this quantity is related to the real-space density field $\delta_{\rm L}(\bm{x})$ as follows:
\begin{align}
\gamma^{\rm I}_{ij}(\bm{x}) &= 
b_{\rm K}
\left[
\mathcal{P}_{ik}(\hat{\bm{x}})\mathcal{P}_{jl}(\hat{\bm{x}})
- \frac{1}{2} \mathcal{P}_{ij}(\hat{\bm{x}})\mathcal{P}_{kl}(\hat{\bm{x}})
\right]
\left( \frac{\partial_{k}\partial_{l}}{\partial^{2}} - \frac{1}{3}\delta_{kl}\right) \delta_{\rm L}(\bm{x})
\notag \\
&= 
b_{\rm K}
\left[
\mathcal{P}_{ik}(\hat{\bm{x}})\mathcal{P}_{jl}(\hat{\bm{x}})
- \frac{1}{2} \mathcal{P}_{ij}(\hat{\bm{x}})\mathcal{P}_{kl}(\hat{\bm{x}})
\right]
\notag \\
& \qquad 
\times \int\frac{{\rm d}^{3}k}{(2\pi)^{3}}
\left( \hat{k}_{k}\hat{k}_{l} - \frac{1}{3}\delta_{kl} \right)
e^{{\rm i}\bm{k}\cdot\bm{x}}\delta_{\rm L}(\bm{k})
~, \label{eq: gamma def}
\end{align}
where we define the projection tensor onto the plane perpendicular to the line-of-sight direction $\mathcal{P}_{ij}(\hat{\bm{x}}) \equiv \delta_{ij} - \hat{x}_{i}\hat{x}_{j}$.
The parameter $b_{\rm K}$ quantifies the response of individual galaxy shapes to the gravitational tidal field.

The projected ellipticity field given in Eq.~(\ref{eq: gamma def}) has two independent degrees of freedom that we denote by $\gamma_{+}$ and $\gamma_{\times}$, defined by
\begin{align}
\left(
\begin{array}{c}
\gamma_{+}(\bm{x}) \\
\gamma_{\times}(\bm{x})
\end{array}
\right)
 &= 
 \left(
\begin{array}{c}
\hat{e}_{1i}(\hat{x})\hat{e}_{1j}(\hat{x}) - \hat{e}_{2i}(\hat{x})\hat{e}_{2j}(\hat{x})\\
2\hat{e}_{1i}(\hat{x})\hat{e}_{2j}(\hat{x})
\end{array}
\right)
\gamma^{\rm I}_{ij}(\bm{x})  \notag \\
& = 
b_{\rm K}
\left(
\begin{array}{c}
\hat{e}_{1i}(\hat{x})\hat{e}_{1j}(\hat{x}) - \hat{e}_{2i}(\hat{x})\hat{e}_{2j}(\hat{x})\\
2\hat{e}_{1i}(\hat{x})\hat{e}_{2j}(\hat{x})
\end{array}
\right)
\notag \\
& \qquad\times 
\int{\frac{{\rm d}^{3}k}{(2\pi)^{3}}}{\rm e}^{{\rm i}\bm{k}\cdot\bm{x}}\, \hat{k}_{i}\hat{k}_{j} \delta_{\rm L}(\bm{k}) ~. \label{eq: gamma field}
\end{align}
In the second equality, we used Eq.~(\ref{eq: gamma def}).
Here we define a two-dimensional orthonormal basis that lies at the tangent space of the sphere at $\bm{x}$, $\hat{\bm{e}}_{1}(\hat{\bm{x}})$ and $\hat{\bm{e}}_{2}(\hat{\bm{x}})$. These vectors satisfy the following relations: $\hat{\bm{x}}\cdot \hat{\bm{e}}_{a} = 0$ and $\hat{\bm{e}}_{a}\cdot \hat{\bm{e}}_{b} = \delta_{ab}$ for $a,b = 1,2$.
The quantities, $\gamma_{+/\times}$, depend on how we specify the two-dimensional orthonormal basis for a given coordinate.
We will give a specific orthonormal basis to compute the GI correlation function in Sec.~\ref{sec: GI corr}.

The parameter, $b_{\rm K}$, is conventionally characterised by using another parameter $A_{\rm IA}$ as~\citep[e.g.,][]{2018JCAP...07..030S,2021MNRAS.501..833K}
\begin{equation}
b_{\rm K} = -0.0134 A_{\rm IA}\Omega_{\rm m,0}/D_{+}(a)~.
\label{eq: AIA to bK}
\end{equation}
The parameter, $A_{\rm IA}$, generally depends on properties of the galaxy/halo populations as well as redshift.
However, the analysis of the simulations shows that this parameter has only a mild redshift dependence for fixed halo properties~\citep{2021MNRAS.501..833K}.
Throughout this paper, we thus assume that the parameter, $A_{\rm IA}$, depends only on halo mass.

We have derived the expression of the projected ellipticity fields in real space.
The ellipticity field $\gamma_{+/\times}$ in the linear alignment model is a first-order quantity of the linear density field, and is not affected by RSD unlike the galaxy density field at the linear level~\citep[see e.g.,][]{2015MNRAS.450.2195S,2020MNRAS.493L.124O,2020MNRAS.494..694O,2021MNRAS.501..833K}. Likewise, relativistic contributions to the ellipticity field are regarded as higher order, and can be ignored. Therefore, we can consider the observed ellipticity field in redshift space to be equivalent to the one in real space, $\gamma_{+/\times}(\bm{s}) \simeq \gamma_{+/\times}(\bm{x})$.

\section{Galaxy and intrinsic alignment correlation}
\label{sec: GI corr}

Having provided the expressions for the density and ellipticity fields, given in Eqs.~(\ref{eq: delta std + delta epsNL})--(\ref{eq: def delta epsNL}) and (\ref{eq: gamma field}), respectively, we present an analytical model of the GI correlation function.
Since the statistical homogeneity and isotropy are not fully ensured in redshift space due to the directional dependence of the observer's line-of-sight, the cross-correlation function between the density $\delta(\bm{s}_1)$ and projected ellipticity $\gamma_{+/\times}(\bm{s}_2)$ fields, which we hereafter assume to be located at the positions $\bm{s}_1$ and $\bm{s}_2$, respectively, is most generally characterized as functions of $\bm{s}_{1}$ and $\bm{s}_{2}$, i.e., $\xi^{\delta\gamma_{+/\times}}(\bm{s}_1,\bm{s}_2)$ (Eq.~(\ref{eq: def GI corr})).
In our analysis, we use a coordinate system such that the unit vectors $\hat{\bm{s}}_{1,2}$ and orthonormal basis $\hat{\bm{e}}_{1,2}$ are given as follows (see below Eq.~(\ref{eq: gamma field}) for the orthonormal condition):
\begin{align}
\hat{\bm{s}}_{1} &= \left( \sin\frac{\theta}{2},\, 0,\, \cos\frac{\theta}{2} \right) ~,~~~
\hat{\bm{s}}_{2} = \left( -\sin\frac{\theta}{2},\,  0,\, \cos\frac{\theta}{2} \right) ~,\notag \\
\hat{\bm{e}}_{1}(\hat{\bm{s}}_{2}) &= 
\left( \cos\frac{\theta}{2}\cos{\varphi},\, -\sin{\varphi},\, \sin\frac{\theta}{2}\cos{\varphi} \right)
~,\notag  \\
\hat{\bm{e}}_{2}(\hat{\bm{s}}_{2}) &=
\left( \cos\frac{\theta}{2}\sin{\varphi},\, \cos{\varphi},\, \sin\frac{\theta}{2}\sin{\varphi} \right)
~,\label{eq: def coordinate}
\end{align}
where we introduce the polar angle $\theta$ to be $\hat{\bm{s}}_{1}\cdot\hat{\bm{s}}_{2} = \cos{\theta}$, and $\varphi$ is defined as the misalignment angle between the unit vector $\hat{\bm e}_2$ and $y$-axis (see left panel in Fig.~\ref{fig: config}).

With the definitions given above, we derive the analytical expression for the GI correlation in Sec.~\ref{sec: analytical xi}.
Since the wide-angle effect causes a significant contribution to the asymmetric correlations, we model the GI correlations without assuming the plane-parallel approximation.
We numerically compute the derived GI correlation functions and present the results in Sec.~\ref{sec: numerical xi}.

\subsection{Analytical model for the GI correlation}
\label{sec: analytical xi}

\begin{figure}
\centering
\includegraphics[width=\columnwidth]{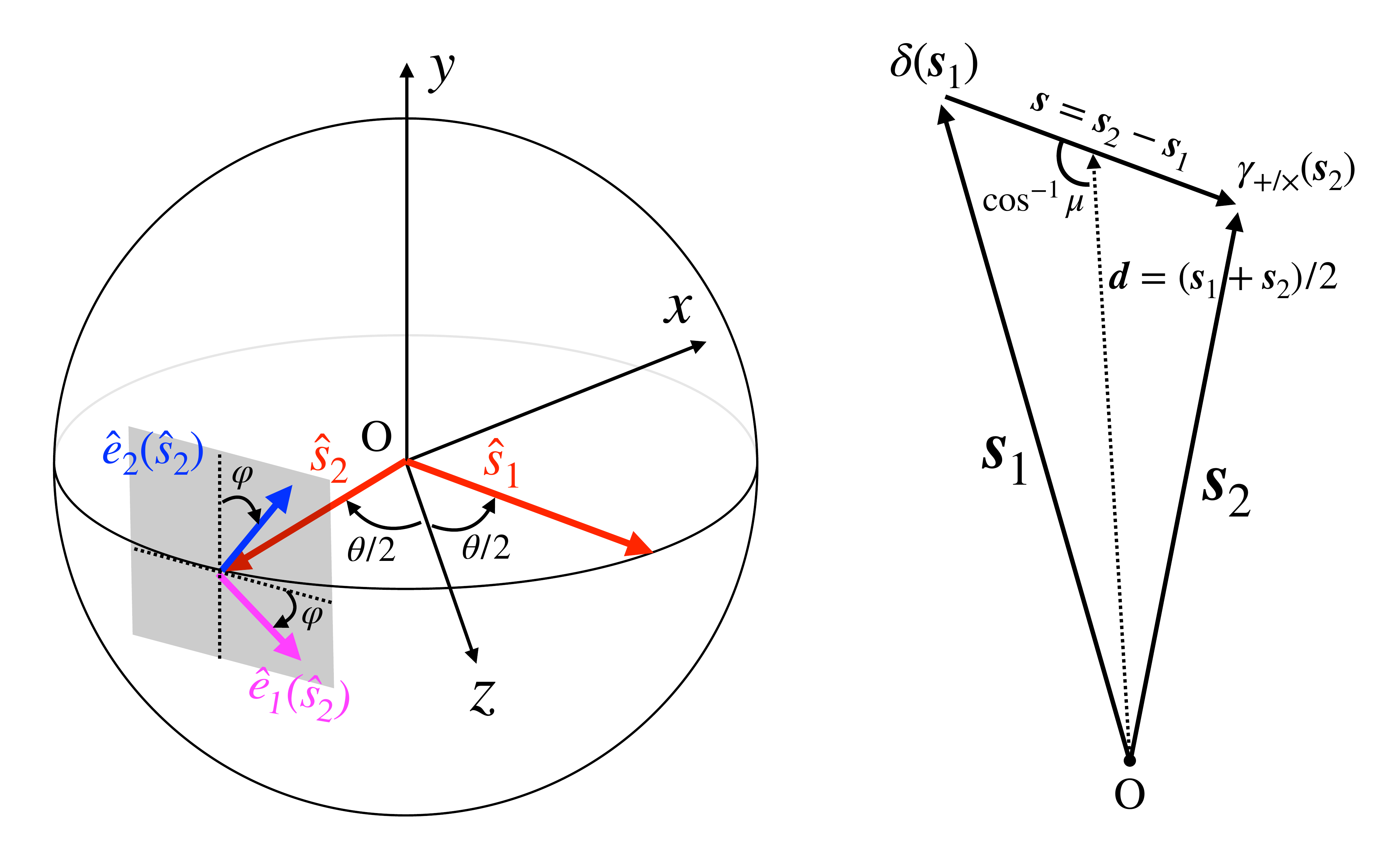}
\caption{
Left: geometric configuration of the unit vectors $\hat{\bm{s}}_{1}$ and $\hat{\bm{s}}_{2}$ pointing respectively to the positions of density and ellipticity fields and orthonormal basis $\hat{\bm{e}}_{1,2}(\hat{\bm{s}}_{2})$.
Both of the vectors lie on the $x$-$z$ plane.
The directional cosine between $\hat{\bm{s}}_{1}$ and $\hat{\bm{s}}_{2}$ is given by $\hat{\bm{s}}_{1}\cdot \hat{\bm{s}}_{2} = \cos{\theta}$.
Right: geometric configuration for the GI correlation. The density $\delta$ and ellipticity field $\gamma_{+/\times}$ are, respectively, observed
at $\bm{s}_1$ and $\bm{s}_2$ with respect to the observer (O).
The separation vector, line-of-sight vector, and directional cosine are defined by $\bm{s} = \bm{s}_{2}-\bm{s}_{1}$, $\bm{d} = (\bm{s}_{1}+\bm{s}_{2})/2$, and $\mu \equiv \hat{\bm{s}}\cdot\hat{\bm{d}}$, respectively.
}
\label{fig: config}
\end{figure}

To compute the redshift-space GI correlation function given in Eq.~(\ref{eq: def GI corr}), we first set $\varphi=0$ in Eq.~(\ref{eq: def coordinate}) without loss of generality (see left panel in Fig.~\ref{fig: config} for the geometrical configuration).
In this configuration, one of the GI correlation functions, $\xi^{\delta\gamma_{\times}}$, always vanishes.
Hereafter, we therefore denote the non-vanishing GI correlation $\xi^{\delta\gamma_{+}}$ by $\xi^{\delta\gamma}$.
Then, following the same steps as in the case of GG correlation and making use of the formulas in Appendix A of \citet{2022MNRAS.511.2732S} (see Eqs.~(A5)--(A9) in their paper), the explicit expressions for the GI correlation function can be derived by substituting Eqs.~(\ref{eq: delta std + delta epsNL}) and (\ref{eq: gamma field}) into Eq.~(\ref{eq: def GI corr}). The result is summarized in the form as
\begin{align}
\xi^{\delta\gamma}(\bm{s}_{1},\bm{s}_{2}) & = 
\sum_{n,\ell} \alpha^{(n)}_{\ell}(\bm{s}_{1},\bm{s}_{2})\, \Xi^{(n)}_{\ell}(s)\sin^{2}\theta
~. \label{eq: xi GI}
\end{align}
Here, we define the function $\Xi^{(n)}_{\ell}$ as
\begin{equation}
\Xi^{(n)}_{\ell}(s) = \int\frac{k^{2}{\rm d}k}{2\pi^{2}}\frac{j_{\ell}(ks)}{(ks)^{n}} P_{\rm L}(k) ~,
\end{equation}
where $j_{\ell}$ and $P_{\rm L}$ are respectively the spherical Bessel function and the linear matter power spectrum defined by $\Braket{\delta_{\rm L}(\bm{k})\delta_{\rm L}(\bm{k}')} = (2\pi)^{3}\delta_{\rm D}(\bm{k}+\bm{k}')P_{\rm L}(k)$.
We summarize the explicit form of the dimensionless coefficients, $\alpha^{(n)}_{\ell}(\bm{s}_{1},\bm{s}_{2})$, in Appendix~\ref{app: full wa}.

The correlation function given in Eq.~(\ref{eq: xi GI}) is invariant under the transformation such that the shape of the triangle, formed with the positions of galaxy and ellipticity fields to cross correlate and the observer, remains unchanged. Thus, instead of using the variables $\bm{s}_{1}$ and $\bm{s}_{2}$ to characterize the GI correlation function, one can describe the GI correlation by introducing the following three variables (see right panel in Fig.~\ref{fig: config}): the separation $s = |\bm{s}| = |\bm{s}_{2}-\bm{s}_{1}|$, line-of-sight distance specifically pointing to a midpoint $d = |\bm{d}| = | \bm{s}_{1} + \bm{s}_{2}|/2$, and directional cosine between the separation vector and the line-of-sight vector $\mu = \hat{\bm{s}}\cdot\hat{\bm{d}}$, i.e., $\xi^{\delta\gamma}(\bm{s}_{1}, \bm{s}_{2}) = \xi^{\delta\gamma}(s, d, \mu)$\footnote{Note importantly that the choice of the line-of-sight vector $\bm{d}$ can affect the analytical expressions for the wide-angle corrections to the GI multipoles. Depending on the choice of its vector, spurious odd multipoles are known to be produced in the GG correlation \citep{2020MNRAS.491.4162T,2016JCAP...01..048R,2017JCAP...01..032G}. The mid-point line-of-sight vector adopted here does not produce such spurious contributions at least at the leading order of the wide-angle correction.}.

We then expand the GI correlation in powers of $(s/d)$ to split the $d$ dependence from the correlation function as follows:
\begin{align}
\xi^{\delta\gamma} (s,d,\mu) &=
\xi^{\delta\gamma}_{\rm pp} (s,\mu)
+ \xi^{\delta\gamma}_{\rm wa} (s,\mu) \left( \frac{s}{d} \right) + O\left( \left( \frac{s}{d} \right)^{2} \right) ~, \label{eq: wa expansion app}
\end{align}
where the first and second terms on the right-hand side are, respectively, the expression in the plane-parallel limit ($s/d\to 0$) and the leading-order correction to the wide-angle effect.

In Eq.~(\ref{eq: wa expansion app}), each of the terms at the right-hand side can be decomposed into four contributions:
\begin{align}
\xi^{\delta\gamma}_{\rm pp/wa}(s,\mu)
& = 
\xi^{(\rm r)}_{\rm pp/wa}(s,\mu)
+ \xi^{(\rm Dop)}_{\rm pp/wa}(s,\mu)
\notag \\
&\qquad \qquad 
+ \xi^{(\rm grav)}_{\rm pp/wa}(s,\mu)
+ \xi^{(\rm nl)}_{\rm pp/wa}(s,\mu) ~, \label{eq: decomposition}
\end{align}
where the first term on the right-hand side represents the real-space GI correlation function. The second, third and fourth terms are the contributions from the Doppler effect, the gravitational redshift due to linear gravitational potential, and that due to the non-perturbative halo potential, respectively.
Then, the four contributions to the plane-parallel limit, i.e., the first term of Eq.~(\ref{eq: wa expansion app}), are given by
\begin{align}
\xi^{(\rm r)}_{\rm pp}(s,\mu) &= 
- b_{\rm K}b(1-\mu^{2}) \Xi^{(0)}_{2}(s) ~, 
\label{eq:xi_r_pp}
\\
\xi^{(\rm Dop)}_{\rm pp}(s,\mu) &= 
- b_{\rm K}f (1-\mu^{2}) 
\frac{1}{7} \left( \Xi^{(0)}_{2}(s) + (1-7\mu^{2})\Xi^{(0)}_{4}(s)\right)
~, \label{eq:xi_Dop_pp} 
\\
\xi^{(\rm grav)}_{\rm pp}(s,\mu) &= - b_{\rm K} \mathcal{M}s \mu (1-\mu^{2}) \Xi^{(1)}_{3}(s) ~, \label{eq: xi grav pp}
\\
\xi^{(\rm nl)}_{\rm pp}(s,\mu) &=
 b_{\rm K}\frac{\phi_{\rm halo}}{aHs} \mu(1-\mu^{2})
\notag 
\\
&
\times\Biggl[
- b \Xi^{(-1)}_{3}(s)
+ \frac{f}{3} \left\{ \left(3 \mu^{2}-1\right) \Xi^{(-1)}_{5}(s) - \Xi^{(-1)}_{3}(s) \right\}
\Biggr]
~, \label{eq: xi eps pp}
\end{align}
and those to the leading-order wide-angle contribution, namely the second term of Eq.~(\ref{eq: wa expansion app}), are given by
\begin{align}
\xi^{(\rm r)}_{\rm wa}(s,\mu) &= 
b_{\rm K}b \mu (1-\mu^{2}) \Xi^{(0)}_{2}(s) ~, 
\label{eq:xi_r_wa}
\\
\xi^{(\rm Dop)}_{\rm wa}(s,\mu) &= 
b_{\rm K}f \mu (1-\mu^{2}) \Xi^{(0)}_{2}(s) ~, 
\label{eq:xi_Dop_wa}
\\
\xi^{(\rm grav)}_{\rm wa}(s,\mu) &= - \frac{b_{\rm K}}{10} \mathcal{M}s (1-\mu^{2}) \left( 8\Xi^{(1)}_{1}(s) + (3-5\mu^{2}) \Xi^{(1)}_{3}(s)\right) ~,
\label{eq:xi_gav_wa}
\\
\xi^{(\rm nl)}_{\rm wa}(s,\mu) &=
\frac{b_{\rm K}}{630}\frac{\phi_{\rm halo}}{aHs} \left(1 - \mu^{2}\right)
\Biggl[
7 \left(5 (9 b-2) \mu^2+9 b-14\right) \Xi^{(-1)}_{3}(s)
\notag \\
& \quad 
-36 (7 b+3) \Xi^{(-1)}_{1}(s) +10 \left(1-7 \mu^2\right) \Xi^{(-1)}_{5}(s)
\notag \\
& \quad
+ f 
\Biggl\{ 7 \left(95 \mu^{2}-17\right) \Xi^{(-1)}_{3}(s)
\notag \\
& \quad 
+5 \left(63 \mu^4-56 \mu^{2}+5\right) \Xi^{(-1)}_{5}(s) -144 \Xi^{(-1)}_{1}(s) \Biggr\}
\Biggr]
~. \label{eq: xi eps wa}
\end{align}
Since the observed shapes of galaxies are projected along the observer's line-of-sight direction, the correlation function becomes anisotropic even in real space and thus has an explicit $\mu$-dependence.
The real-space and Doppler contributions contain only even powers of $\mu$ in the plane-parallel limit and only odd powers of $\mu$ in the wide-angle correction, while the gravitational redshift contributions have the opposite $\mu$-dependence. 
This result suggests that the gravitational redshift effects cause qualitatively different anisotropy in the GI correlations than the other two contributions.
Note that the expressions of the real-space and Doppler terms in the plane-parallel limit, Eqs.~(\ref{eq:xi_r_pp}) and (\ref{eq:xi_Dop_pp}), exactly coincide with those presented in \citet{2020MNRAS.493L.124O}.

To quantify the anisotropies of the GI correlation, we use the multipole expansion of the correlation function.
We define the multipole moments by taking an average of the correlation function over the directional cosine $\mu$, weighting with the Legendre polynomials $\mathcal{L}_{\ell}(\mu)$:
\begin{equation}
\xi^{\delta\gamma}_{\ell}(s)
= \frac{2\ell+1}{2}\int^{1}_{-1}{\rm d}\mu\; \xi^{\delta\gamma}(s,\mu) \mathcal{L}_{\ell}(\mu)~.
\end{equation}
We summarize all non-vanishing multipoles in Appendix~\ref{app: multipoles}. Note interestingly that the sum of these multipole moments is shown to vanish in each contribution, and we have 
\begin{equation}
\sum_{\ell}\xi^{(\rm x)}_{{\rm pp/wa}, \ell}(s)=0~, \label{eq: moment sum}
\end{equation}
where ${\rm x}=$ \{r, Dop, grav, nl\}. As we discussed in Appendix~\ref{app: multipoles}, these relations come from the fact that the GI correlation function involves the factor of $(1-\mu^2)$, arising from projection of the ellipticity field (see Eqs.~(\ref{eq:xi_r_pp})--(\ref{eq: xi eps wa}).

As a highlight of the results, we here present the expressions of the dipole moment, which will be used to estimate the signal-to-noise ratio in next section:
\begin{align}
\xi^{({\rm r})}_{\rm wa,1}(s) &= \frac{2}{5}\,b_{\rm K}b \Xi^{(0)}_{2}(s)
~, \label{eq: dipole real}\\
\xi^{({\rm Dop})}_{\rm wa,1}(s) &= \frac{2}{5}\,b_{\rm K}f \Xi^{(0)}_{2}(s)
~, \label{eq: dipole std}\\
\xi^{({\rm grav})}_{\rm pp,1}(s) &= - \frac{2}{5}\mathcal{M}sb_{\rm K}\Xi^{(1)}_{3}(s) ~, \label{eq: dipole grav}\\
\xi^{(\rm nl)}_{\rm pp,1}(s) &= - \frac{2}{5}\frac{\phi_{\rm halo}}{aHs}b_{\rm K}b\,\Xi^{(-1)}_{3}(s)
\nonumber
\\
&- \frac{2}{105}f\frac{\phi_{\rm halo}}{aHs}b_{\rm K}\left( 7\Xi^{(-1)}_{3}(s) - 2\Xi^{(-1)}_{5}(s)\right)
~.\label{eq: dipole eps}
\end{align}
In the above, the non-vanishing real-space and Doppler contributions arise from the wide-angle correction, and their scale-dependence is described by the same function $\Xi_2^{(0)}$. That is, these are proportional to each other. Note interestingly that 
a similar relation is also obtained from the octupole moment. Since the higher-odd multipoles of $\ell\geq5$ vanish at the order of $\mathcal{O}(s/d)$, 
Eq.~(\ref{eq: moment sum}) leads to 
\begin{align}
\xi^{({\rm r})}_{\rm wa,1}(s) = - \xi^{({\rm r})}_{\rm wa,3}(s) = (b/f)\,\xi^{({\rm Dop})}_{\rm wa,1}(s) = -(b/f)\,\xi^{({\rm Dop})}_{\rm wa,3}(s) ~.
\label{eq:dipole_octupole_xi_wa}
\end{align}

On the other hand, the gravitational redshift contributions, $\xi^{{\rm(grav)}}$ and $\xi^{{\rm(nl)}}$, appear non-vanishing in the plane-parallel limit at the leading order. Since Eq.~(\ref{eq: moment sum}) holds for each of the gravitational redshift contributions, one can also derive expressions similar to 
Eq.~(\ref{eq:dipole_octupole_xi_wa}).  An explicit example is $\xi_{\rm pp,3}^{{\rm (grav)}}=-\xi_{\rm pp,1}^{{\rm (grav)}}$. 
This indicates that the amplitudes of the dipole and higher multipole moments are comparable to each other. 
Thus, compared to those arising from the wide-angle effect, we expect the gravitational redshift contributions, including the higher multipoles, to be dominant at the scales sufficiently smaller than the line-of-sight distance, i.e., $s/d \ll 1$. By explicitly computing each contribution of the GI multipoles, we will see below that the properties and relations discussed here indeed hold (see Fig.~\ref{fig: xi ell}).

\subsection{Numerical results}
\label{sec: numerical xi}

We now present the multipole moments of the GI correlation functions by using the expressions derived in the previous subsection.
In computing the GI correlation, we need to specify the bias parameters of the density and ellipticity fields, $b$ and $b_{\rm K}$, respectively.
For the former, we adopt the halo model prescription with the mass function given by \citet{1999MNRAS.308..119S}. Then, the density bias parameter, $b$, is uniquely determined for a given halo mass. 
To determine $b_{\rm K}$, we use the relation Eq.~(\ref{eq: AIA to bK}) and the fact that the parameter $A_{\rm IA}$ depends very weakly on redshift for the fixed halo mass. 
Following the simulation result of \cite{2021MNRAS.501..833K}, we adopt the values as $A_{\rm IA} = 15$, $23$, and $28$ for the halo masses $M/(M_{\odot}/h) = 10^{12}$, $10^{13}$, and $10^{14}$, respectively.
Hence, provided the linear power spectrum, redshift and halo mass, we can compute the GI correlation function.
Since the amplitude of the GI correlation in our model is linearly proportional to the strength of the IAs, $b_{\rm K}$, we show below the resultant multipoles divided by $b_{\rm K}$, i.e., $\xi^{\delta\gamma}_{\ell}(s)/b_{\rm K}$.

\begin{figure*}
\centering
\includegraphics[width=0.9\textwidth]{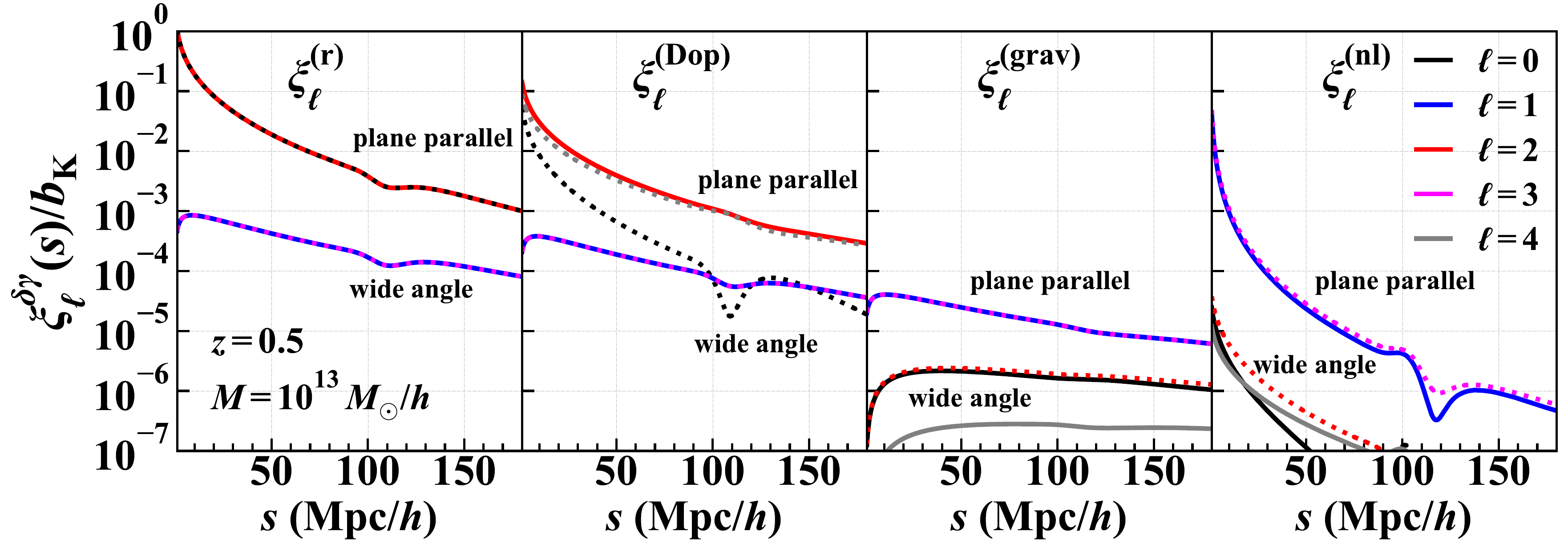}
\caption{
Multipole moments of GI correlation functions for halo mass $M=10^{13}\, M_{\odot}/h$ at $z=0.5$ divided by the strength of IAs $b_{\rm K}$, which is a negative quantity.
The multipoles are shown up to $\ell = 4$.
From left to right panels, we show the results of the real-space, Doppler, linear gravitational redshift, and non-linear gravitational redshift terms, respectively.
The dotted lines indicate the negative values.
}
\label{fig: xi ell}
\end{figure*}

In Fig.~\ref{fig: xi ell}, the multipole moments of the GI correlation functions computed up to $\ell=4$ are shown, setting specifically the halo mass and redshift to $M=10^{13}\,M_{\odot}/h$ and $z=0.5$, respectively.
From left to right, we separately plot the real-space, Doppler, linear gravitational redshift, and non-linear gravitational redshift contributions.
Among the four contributions, the real-space and Doppler contributions can produce the largest amplitudes in the GI correlation, as we expected. In particular, the even multipoles arising from the plane-parallel limit dominate their contributions. A closer look at their amplitudes reveals that the real-space contribution is larger than the Doppler contribution. This is due to the fact that we have $b/f>1$ in the current setup. On the other hand, the gravitational redshift contributions, i.e., $\xi_\ell^{{\rm(grav)}}$ and $\xi_\ell^{{\rm(nl)}}$, have amplitudes smaller than the real-space and Doppler correlations, and may be regarded as sub-dominant contributions. Nevertheless, their leading contributions appear at the odd multipoles, and the dipole and octupole moments have the same or comparable amplitudes, consistently with what we discussed in Sec.~\ref{sec: analytical xi}. A notable point may be that the dipole and octupole amplitudes of $\xi_\ell^{{\rm(nl)}}$ get increased significantly as we go to small scales. 
This suggests that the odd multipoles of the small-scale GI correlation can be a sensible probe of the gravitational redshift effect especially induced by the halo potential.
Overall, the wide-angle corrections, which appear at odd multipoles for $\xi_\ell^{{\rm(r)}}$ and $\xi_\ell^{{\rm(Dop)}}$ and even multipoles for $\xi_\ell^{{\rm(grav)}}$ and $\xi_\ell^{{\rm(nl)}}$, are all subdominant. While the former contributions follow the relation at Eq.~(\ref{eq:dipole_octupole_xi_wa}), all of their amplitudes are smaller than the predicted amplitudes in the plane-parallel limit.

\begin{figure*}
\centering
\includegraphics[width=0.9\textwidth]{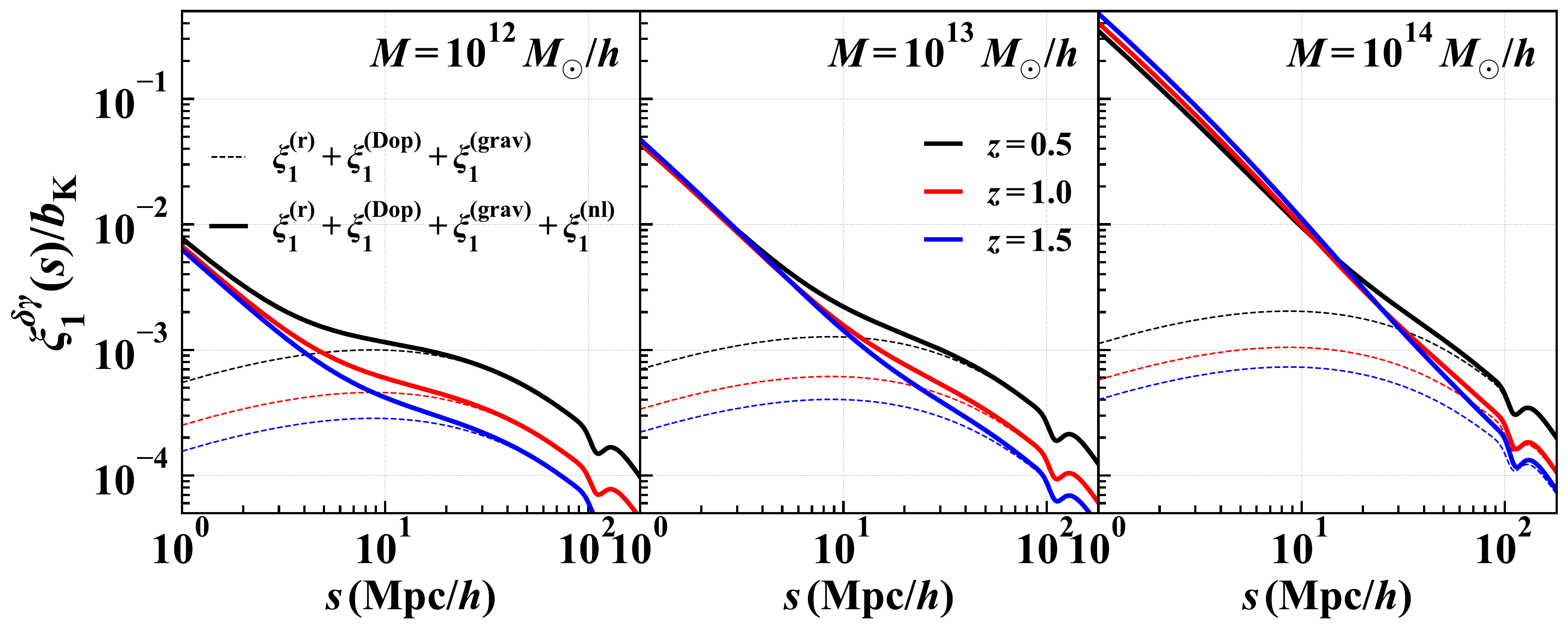}
\caption{Dipole moments of GI correlation function at various redshifts with halo masses $M/(M_{\odot}/h)=10^{12}$ (left), $10^{13}$ (center), and $10^{14}$ (right).
The solid and dotted lines represent the analytical predictions including and neglecting the non-linear gravitational redshift contributions, respectively.
As in Fig.~\ref{fig: xi ell}, we present the dipole moments divided by the strength of the IAs, $\xi_{1}/b_{\rm K}$.
}
\label{fig: xi 1}
\end{figure*}

In Fig.~\ref{fig: xi 1}, to discuss the detectability of the relativistic effect in next section, we focus on the lowest-order odd multipole, i.e,. GI dipole, and show its dependence on the halo mass and redshift.
From the left to right panels, we present the predictions of the GI dipole for the three different host halo masses, $M=10^{12}$, $10^{13}$, and $10^{14}\,M_{\odot}/h$, respectively (from left to right panels). In each panel we also show the results for different redshifts,
$z=0.5$ (black), $1.0$ (red), and $1.5$ (blue)
The solid and dotted lines represent the dipole predictions including and neglecting the halo potential, respectively.
One can see that the non-perturbative potential term significantly enhances the amplitude of the dipole correlation on small scales ($s\lesssim 10\, {\rm Mpc}/h$), as expected from Fig.~\ref{fig: xi ell}.
This effect becomes even more prominent when the potential deepens namely, when massive haloes or higher redshifts are considered.

Here, we summarise the notable differences between behaviours of the dipoles of GI and GG correlations~\citep{2020MNRAS.498..981S,2022MNRAS.511.2732S}.
First, the GG dipole exhibits a sign flip at small scales because the gravitational redshift effect flips the relative positions of the two different biased objects in redshift space.
The GI dipole does not show such a sign flip.
In our model of the GI correlation, the ellipticity field is hardly affected by RSD, but the position of density fields in redshift space appears to shift away from the observer due to the gravitational redshift effect.
Second, while we need two populations of galaxies to detect the gravitational redshift effect in the GG correlation, we can measure the signal in the GI correlation using a single galaxy population with the galaxy shape information.
It is an advantage of considering the GI correlation as a probe of the gravitational redshift effect, complimentary to the GG correlation.

\section{Detectability of the GI dipole}
\label{sec: detectability}

Having provided the analytical model, in this section, we discuss the detectability of the GI correlation.
The gravitational redshift effect can give the most significant contribution to odd multipoles, of which the dipole is expected to have a larger signal-to-noise ratio. Hence, we hereafter focus on the dipole, and estimate the signal-to-noise ratio. The detectability of the GI octupole would be an interesting issue to be discussed, but we leave it to our future work. 
We first present the covariance matrix for the GI dipole correlation in Sec.~\ref{sec: covariance}, with a detailed derivation given in Appendix~\ref{app: covariance}. After showing quantitative results for the covariance matrix in Sec.~\ref{sec: covariance numerical}, we compute the 
signal-to-noise ratio for the GI dipole in several representative setups in Sec.~\ref{sec: SN}.

\subsection{Covariance matrix for GI dipole}
\label{sec: covariance}

While statistical quantities are in principle obtained by an ensemble average, they are estimated by a volume integral in cosmological observations. This finite-volume effect can induce some fluctuations, which lead to a non-negligible correlation between different scales in the measured quantities. This is true for the correlation function even if the observed density and/or ellipticity fields follow the Gaussian statistics. Here, we present an analytical expression of the Gaussian contribution to the covariance matrix for the GI dipole, based on the estimator defined below.

Denoting the two fields by $X$ and $Y$ where $\{XY\}=\{ \delta\delta,\delta\gamma,\gamma\gamma\}$, we have as the estimator for the dipole moments of the correlation functions, $\hat{\xi}^{\rm XY}_{1}$,\footnote{The hatted quantities, $\hat{\xi}_1^{\rm XY}$ and $\hat{\xi}^{\rm XY}$, which specifically appear in Sec.~\ref{sec: detectability} and Appendix \ref{app: covariance}, imply an estimator of the GI correlation, and should not be confused with a unit vector.}
\begin{align}
\hat{\xi}^{\rm XY}_{1}(s) = \frac{3}{2}\int^{1}_{-1}{\rm d}\mu\,\mu\,\int\frac{{\rm d}^{3}r}{V}\, {\rm X}\left(\bm{r}-\frac{\bm{s}}{2}\right){\rm Y}\left(\bm{r}+\frac{\bm{s}}{2}\right) ~, \label{eq: def estimator}
\end{align}
where $V$ and $\mu$ respectively stand for the observed volume and the directional cosine between the (fixed) line-of-sight $\hat{\bm{z}}$ and separation vectors defined by $\mu \equiv \hat{\bm{s}}\cdot\hat{\bm{z}}$.

Provided the definition of the estimator above, we analytically compute the covariance matrix of the GI dipole, which is given by
\begin{align}
{\rm COV}_{1}(\bm{s},\bm{s}')
\equiv
\Braket{\hat{\xi}^{\delta\gamma}_{1}(\bm{s})\hat{\xi}^{\delta\gamma}_{1}(\bm{s}')} - \Braket{\hat{\xi}^{\delta\gamma}_{1}(\bm{s})} \Braket{\hat{\xi}^{\delta\gamma}_{1}(\bm{s}')} ~. \label{eq: cov 1st def}
\end{align}
Based on Eq.~(\ref{eq: cov 1st def}), the covariance matrix for the GI dipole is derived in Appendix~\ref{app: covariance}. The resultant expression is given by
\begin{align}
& {\rm COV}_{1}(s,s')
\notag \\
&
= 
\frac{1}{V}
\int\frac{k^{2}{\rm d}k}{2\pi^{2}}
\sum_{L,L'}\sum_{M,M'}
j_{L}(ks)j_{L'}(ks')
\notag \\
& \quad
\times \sum_{\ell_{1},m_{1}}
\sum_{\ell_{2},m_{2}}
\Bigl[ P^{\delta\delta}_{\ell_{1},m_{1}}(k)P^{\gamma\gamma}_{\ell_{2},m_{2}}(k)
+ (-1)^{L'}P^{\delta\gamma}_{\ell_{1},m_{1}}(k)P^{\delta\gamma}_{\ell_{2},m_{2}}(k)\Bigr]
 \notag \\
& \quad
\times
\mathcal{G}_{L,M,L',M'}^{(\ell_{1},m_{1})(\ell_{2},m_{2})}
\mathcal{I}_{1,L,M}
\mathcal{I}_{1,L',M'}
\notag \\
& + \frac{1}{V}\int\frac{k^{2}{\rm d}k}{2\pi^{2}}
\sum_{L,L'}\sum_{M,M'}
j_{L}(ks)j_{L'}(ks')
\notag \\
& \quad 
\times
\sum_{\ell_{3},m_{3}}
\left[ P^{\delta\delta}_{\ell_{3},m_{3}}(k) N^{\gamma} + P^{\gamma\gamma}_{\ell_{3},m_{3}}(k) N^{\delta} \right]
\mathcal{G}_{L,M,L',M'}^{(\ell_{3},m_{3})}
\mathcal{I}_{1,L,M}
\mathcal{I}_{1,L',M'}
\notag \\
& + \frac{3N^{\delta}N^{\gamma} }{4\pi s^{2} L_{\rm p} V} \delta^{\rm K}_{s,s'} ~,
\label{eq: cov dipole}
\end{align}
where $\delta^{\rm K}_{\rm X,Y}$ and $L_{\rm p}$, respectively, represent the Kronecker delta and the side-length of square pixels, which we fix to $2\, {\rm Mpc}/h$ in this paper~\citep{2018JCAP...05..043L,2022MNRAS.511.2732S}.
The quantities $\mathcal{G}_{L,M,L',M'}^{(\ell_{1},m_{1})(\ell_{2},m_{2})}$, $\mathcal{G}_{L,M,L',M'}^{(\ell_{3},m_{3})}$, and $\mathcal{I}_{\ell,L,M}$ are the real-valued functions expressed in terms of the Wigner 3-j symbols, whose explicit forms are given in Appendix~\ref{app: covariance} (see Eqs.~(\ref{eq: def calG l1l2}), (\ref{eq: def calG l3}), and (\ref{eq: def calI}), respectively).
The noise contribution, $N^{\rm X}$, indicates $N^{\delta} = 1/n_{\rm g}$ for the galaxy density field ($X=\delta$) or $N^{\gamma} = \sigma^{2}_{\rm shape}/n_{\rm g}$ for the ellipticity field ($X=\gamma$), where $n_{\rm g}$ and $\sigma_{\rm shape}$ are the galaxy number density and rms of the galaxy's ellipticity, respectively.
The functions $P^{\rm XY}_{\ell,m}(k)$ stand for the spherical harmonic coefficients of the Fourier counterpart of the correlation function:
\begin{align}
\xi^{\rm XY}(\bm{s})  = \int\frac{{\rm d}^3k}{(2\pi)^{3}}e^{-i\bm{k\cdot\bm{s}}} \sum_{\ell, m}P^{\rm XY}_{\ell,m}(k)Y_{\ell,m}(\hat{\bm{k}}) ~, \label{eq: xi to Pelllm}
\end{align}
with $Y_{\ell,m}(\hat{\bm{k}})$ being the spherical harmonics.
The explicit expressions for the coefficients $P^{\rm XY}_{\ell,m}$ in the plane-parallel limit are summarized in Appendix~\ref{app: coefficients Ylm}. 
Note that we adopt the plane-parallel approximation in Eq.~(\ref{eq: xi to Pelllm}), with which the correlation function becomes a function of the separation vector only. This approximation is used only for the estimation of the covariance matrix and known to induce negligibly small biases to the resulting covariance matrix \citep[e.g.,][]{2018JCAP...05..043L}.

While the first and third terms in the right-hand side of Eq.~(\ref{eq: cov dipole}) respectively represent the contributions arising purely from the cosmic variance (CV$\times$CV) and Poisson noise (P$\times$P), the second term describes the cross talk between them (CV$\times$P). 
Note that in principle the summation with respect to the multipoles $L$ and $L'$ has to be taken from $0$ to infinity. However, we numerically confirmed that contributions from the higher multipoles decrease as increasing $L$ and $L'$. In the analysis presented below, we truncate the summation at $L_{\rm max}=20$, which still gives an accurate estimation of the covariance at a sub-percent level for our interest range of separation, $1 \,{\rm Mpc}/h\, <s,\,s'<100 \,{\rm Mpc}/h$.

\subsection{Numerical results of the GI covariance}
\label{sec: covariance numerical}

In this subsection, we compute Eq.~(\ref{eq: cov dipole}) numerically and investigate its quantitative behaviour, focusing particularly on the diagonal components. For this purpose we need to specify the number density of galaxies $n_{\rm g}$ and the root mean square of the ellipticity $\sigma_{\rm shape}$. In what follows, we adopt the Sheth-Tormen mass function~\citep{1999MNRAS.308..119S} to determine $n_{\rm g}$ and set $\sigma_{\rm shape}$ to a typical value for upcoming galaxy surveys, $\sigma_{\rm shape}=0.2$.

\begin{figure}
\centering
\includegraphics[width=0.9\columnwidth]{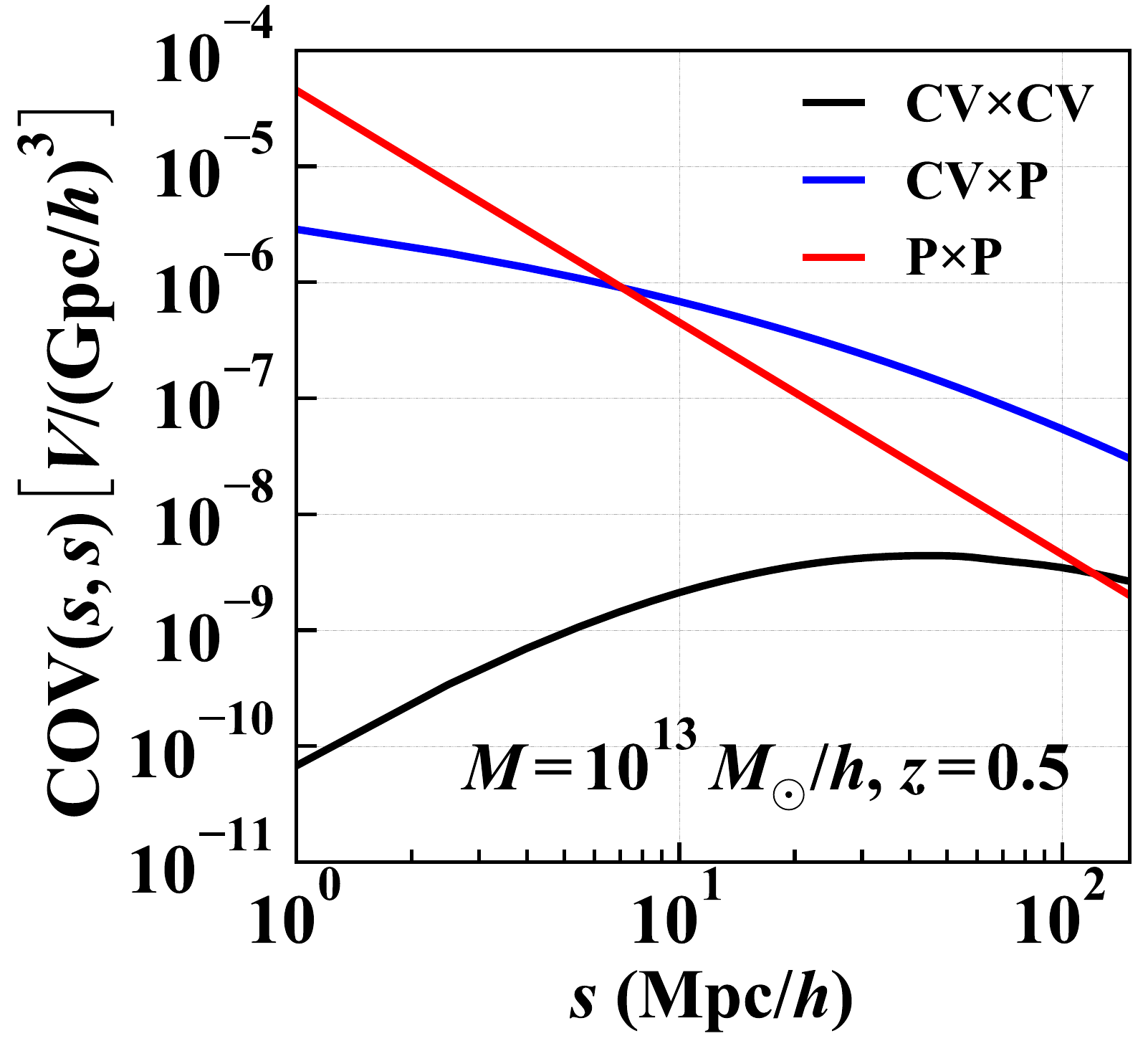}
\caption{Diagonal components of the covariance matrix multiplied by the observed volume as a function of separation.
The black, blue, and red solid lines represent the contribution to the covariance matrix from the CV$\times$CV, CV$\times$P, and P$\times$P terms, respectively, where CV stands for the cosmic variance and P for the Poisson term.
The parameters used in these calculations are as follows: $z=0.5$, $M=10^{13}\, M_{\odot}/h$, $\sigma_{\rm shape}=0.2$, and $L_{\rm p} = 2\, {\rm Mpc}/h$.
We use the parameter $A_{\rm IA} = 23$ taken from the simulations.}
\label{fig: cov}
\end{figure}

Fig.~\ref{fig: cov} plots separately the three contributions to the diagonal components of the covariance matrix, i.e., CV$\times$CV (black), CV$\times$P (blue) and P$\times$P (red), divided by the survey volume.
We show the results only for the halo mass of $10^{13}\,M_\odot/h$ at $z=0.5$. We find that the dominant contributions to the covariance matrix come from the CV$\times$P and P$\times$P terms. The cosmic variance term, CV$\times$CV, can be ignored unless we consider large separations, $s\gtrsim80 \,{\rm Mpc}/h$. The reason why the CV$\times$CV term is small is attributed to a partial cancellation of the summation in Eq.~(\ref{eq: cov dipole}):
\begin{align}
& \sum_{\ell_{1},m_{1}}
\sum_{\ell_{2},m_{2}}
\mathcal{G}_{1,0,1,0}^{(\ell_{1},m_{1})(\ell_{2},m_{2})}
\Bigl[ P^{\delta\delta}_{\ell_{1},m_{1}}(k)P^{\gamma\gamma}_{\ell_{2},m_{2}}(k)
\notag \\
&
-
\left( P^{\delta\gamma({\rm r})}_{\ell_{1},m_{1}}(k) + P^{\delta\gamma({\rm Dop})}_{\ell_{1},m_{1}}(k) \right)
\left( P^{\delta\gamma({\rm r})}_{\ell_{2},m_{2}}(k) + P^{\delta\gamma({\rm Dop})}_{\ell_{2},m_{2}}(k)\right)
\Bigr]
= 0 ~.
\end{align}
Note that a similar cancellation has been seen in the covariance of the GG dipole~\citep{2016JCAP...08..021B,2017PhRvD..95d3530H,2022MNRAS.511.2732S}.

\begin{figure}
\centering
\includegraphics[width=0.9\columnwidth]{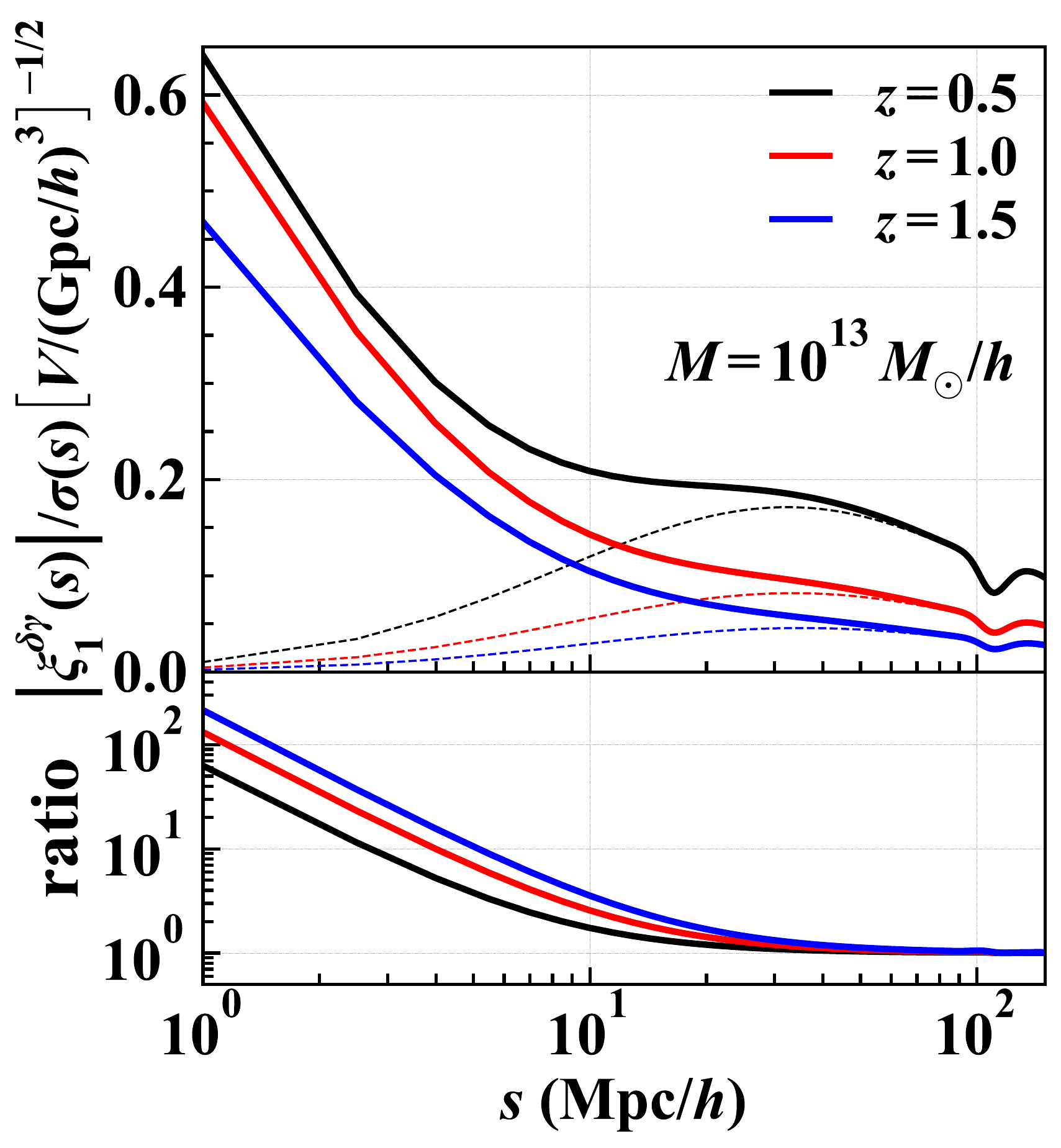}
\caption{
Top: dipole moment divided by the square root of the diagonal components of the covariance matrix multiplied by observed volume, $\left| \xi^{\delta\gamma}_{1}(s) \right|/\sigma(s) V^{-1/2}$, where $\sigma(s)$ is the square root of the diagonal components of the covariance matrix, $\sigma(s) \equiv \left( {\rm COV}(s,s) \right)^{1/2}$, for various redshift with halo mass $M=10^{13}\, M_{\odot}/h$.
The solid and dashed lines represent the result with and without the non-perturbative halo potential contributions, respectively.
Bottom: the ratio between the results including and ignoring the halo potential term.
All the parameters used in this plot are the same as Fig.~\ref{fig: cov}.
}
\label{fig: xi1 over cov}
\end{figure}

he top panel of Fig.~\ref{fig: xi1 over cov} shows the ratio of GI dipole to its expected error, $\left| \xi^{\delta\gamma}_{1}(s) \right|/\sigma(s)$, where $\sigma(s)$ is the square root of the diagonal covariance $\sigma(s)\equiv \left({\rm COV}(s,s)\right)^{1/2}$, fixing again the mass of haloes to $10^{13}\,M_\odot/h$. In plotting the results, we further dividing the ratio by square root of the survey volume, and plot it as a function of separation at $z=0.5$ (black), and $1$ (red) and $1.5$ (blue).
Supposing that the off-diagonal components of the covariance matrix is small, the ratio $\left| \xi^{\delta\gamma}_{1}(s) \right|/\sigma(s)$ approximately describes the signal-to-noise ratio of the GI dipole at each separation $s$. The solid and dashed lines represent the results with and without the gravitational redshift effect induced by the halo potential, respectively. To clarify the difference between the two curves, their ratio is shown in the bottom panel of Fig.~\ref{fig: xi1 over cov}. This demonstrates that the gravitational redshift from halo potential contributes significantly to the total signal-to-noise ratio, and it becomes more prominent as decreasing the separation. Without the halo potential term, the small-scale contribution is rather suppressed. As a result, the ratio shown in the bottom panel is dramatically enhanced. 
Although Fig.~\ref{fig: xi1 over cov} reveals a part of the contributions to the signal-to-noise ratio, the trend is shown to remain unchanged when we estimate the total signal-to-noise ratio in next subsection.

\subsection{Signal-to-noise ratio}
\label{sec: SN}

We are in position to quantitatively discuss the detectability of the GI dipole by explicitly computing the {\it cumulative} signal-to-noise ratio below:
\begin{align}
\left( \frac{\rm S}{\rm N}\right)^{2}
& =
\sum^{s_{\rm max}}_{s_{1},s_{2} = s_{\rm min}}
\xi^{\delta\gamma}_{1}(s_{1})\, \left( {\rm COV}_{1}(s_{1},s_{2}) \right)^{-1}\, \xi^{\delta\gamma}_{1}(s_{2})
~, \label{eq: def SN}
\end{align}
where the parameters $s_{\rm min}$ and $s_{\rm max}$ respectively stand for the minimum and maximum separations.
We set $s_{\rm max}$ to $150\, {\rm Mpc}/h$, below which the systematic effect due to the plane-parallel limit assumed for the estimation of the covariance is negligible~\citep[see e.g.,][]{2017PhRvD..95d3530H}.
As for $s_{\rm min}$, our model does not take into account baryonic effects, which can affect the GI correlation on small scales. We thus restrict our estimation of the signal-to-noise to the scales where such an effect can be neglected, and set $s_{\rm min} \geq 1\, {\rm Mpc}/h$.
Although this choice might be optimistic,
our primary goal here is to clarify the potential power to detect relativistic effects from the GI dipole. Hence, we shall take $s_{\rm min}$ to be aggressively small. 
In the next paragraph we examine the dependence of the signal-to-noise ratio on $s_{\rm min}$ values.
Since the signal-to-noise ratio is proportional to the square root of the observed volume, the estimated signal-to-noise ratios are all normalized by $V^{-1/2}$.

\begin{figure}
\centering
\includegraphics[width=\columnwidth]{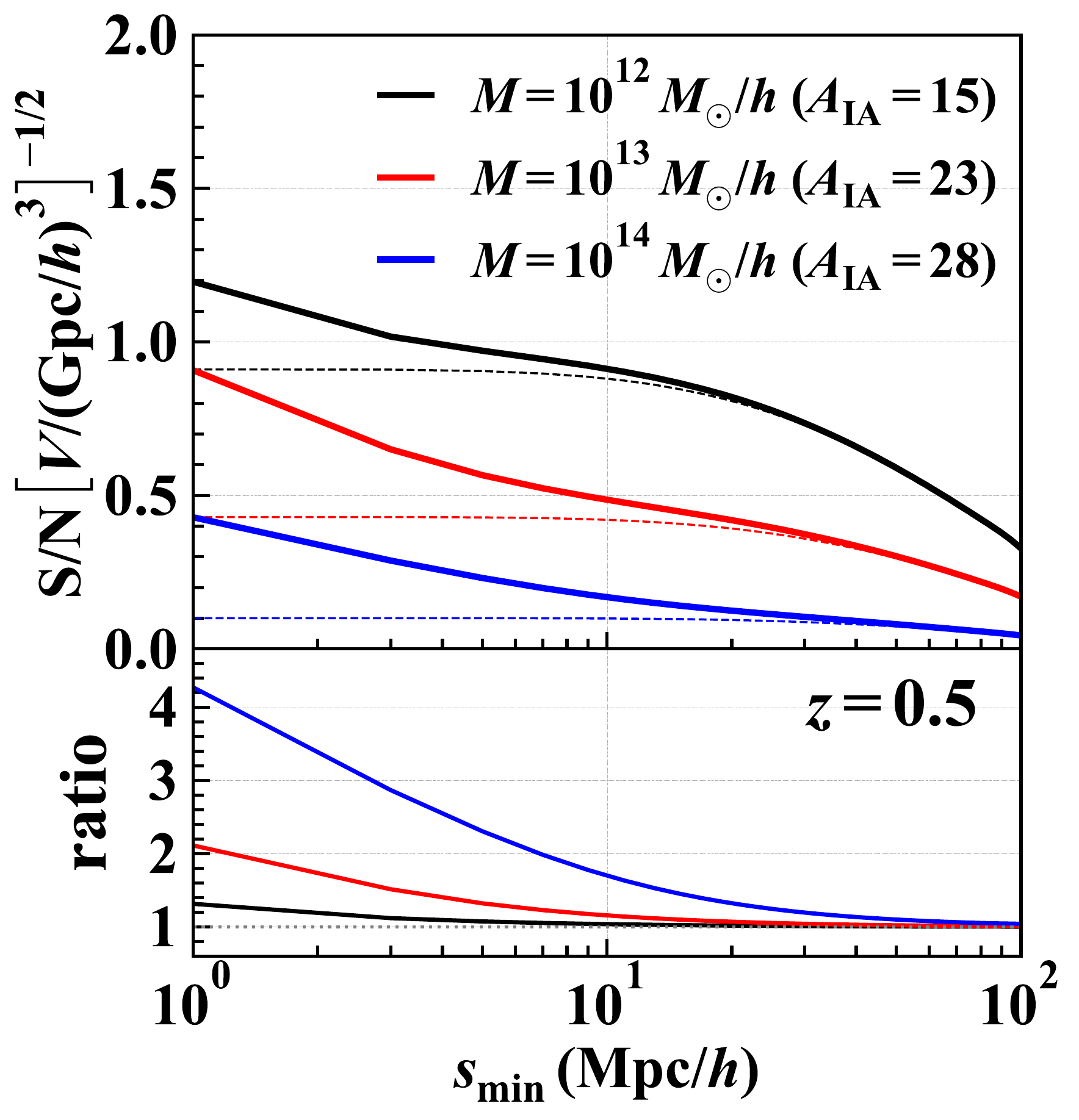}
\caption{Top: signal-to-noise ratio normalized by the square root of the observed volume as a function of the minimum separation $s_{\rm min}$ at $z=0.5$.
The solid and dashed lines in the top panel show the result with and without the non-perturbative potential contributions, respectively.
Bottom: ratio of the signal-to-noise ratios between with and without the non-perturbative potential contribution, $\left({\rm S/N}\right)/\left({\rm S/N} (\phi_{\rm halo} = 0) \right)$.
The horizontal dotted line is equal to unity.
The maximum separations are fixed to $s_{\rm max} = 150\, {\rm Mpc}/h$, and the other parameters are the same as Fig.~\ref{fig: cov}.
As indicated in the figure, we use the parameter $A_{\rm IA} = 15$, $23$, and $28$ for $M=10^{12}$, $10^{13}$, and $10^{14}\, M_{\odot}/h$, respectively, taken from \citet{2021MNRAS.501..833K} (see also Eq.~(\ref{eq: AIA to bK})).
}
\label{fig: SN smin}
\end{figure}

First we show the signal-to-noise ratio as a function of the minimum separation $s_{\rm min}$ in Fig.~\ref{fig: SN smin}, fixing the redshift to $z=0.5$. The results for the halo masses of $M/(M_{\odot}/h)=10^{12}$ (black), $10^{13}$ (red), and $10^{14}$ (blue) are plotted, adopting the IA amplitudes $A_{\rm IA}=15$, $23$, and $28$, respectively~\citep{2021MNRAS.501..833K}. The other parameters are set to the values shown in Fig.~\ref{fig: cov}.
In the top panel, the solid and dashed lines respectively show the signal-to-noise ratios with and without the non-perturbative potential contribution.
Including the gravitational redshift from the halo potential, the signal-to-noise ratio exhibits a strong enhancement at $s_{\rm min}\lesssim 10\, {\rm Mpc}$, and deviates from the results ignoring the halo potential.
These behaviours are consistent with the results in Fig.~\ref{fig: xi1 over cov}.
Another notable point is that the signal-to-noise ratio increases as decreasing the halo mass.
The reason for this comes from the fact that the shot noise contributions, i.e., ${\rm CV}\times {\rm P}$ and ${\rm P}\times {\rm P}$ terms, dominate the covariance matrix. 
Thus, since a low-mass halo is more abundant than a high-mass one, the former is expected to have a better signal-to-noise ratio.

In the lower panel of Fig.~\ref{fig: SN smin}, we estimate how the gravitational redshift effect from the halo potential improves the detectability of the GI dipole, by taking the ratio of the signal-to-noise ratios with and without the non-perturbative potential shown in the upper panel.
As increasing the halo mass, the impact of the halo potential term on the signal-to-noise ratio becomes more significant, but at the same time, the signal-to-noise ratio itself gets smaller, as shown in the upper panel.
This behaviour is again consistent with the results in Fig.~\ref{fig: xi1 over cov}, indicating that one needs a careful sample selection for a practical measurement of the gravitational redshift effect.

\begin{figure}
\centering
\includegraphics[width=\columnwidth]{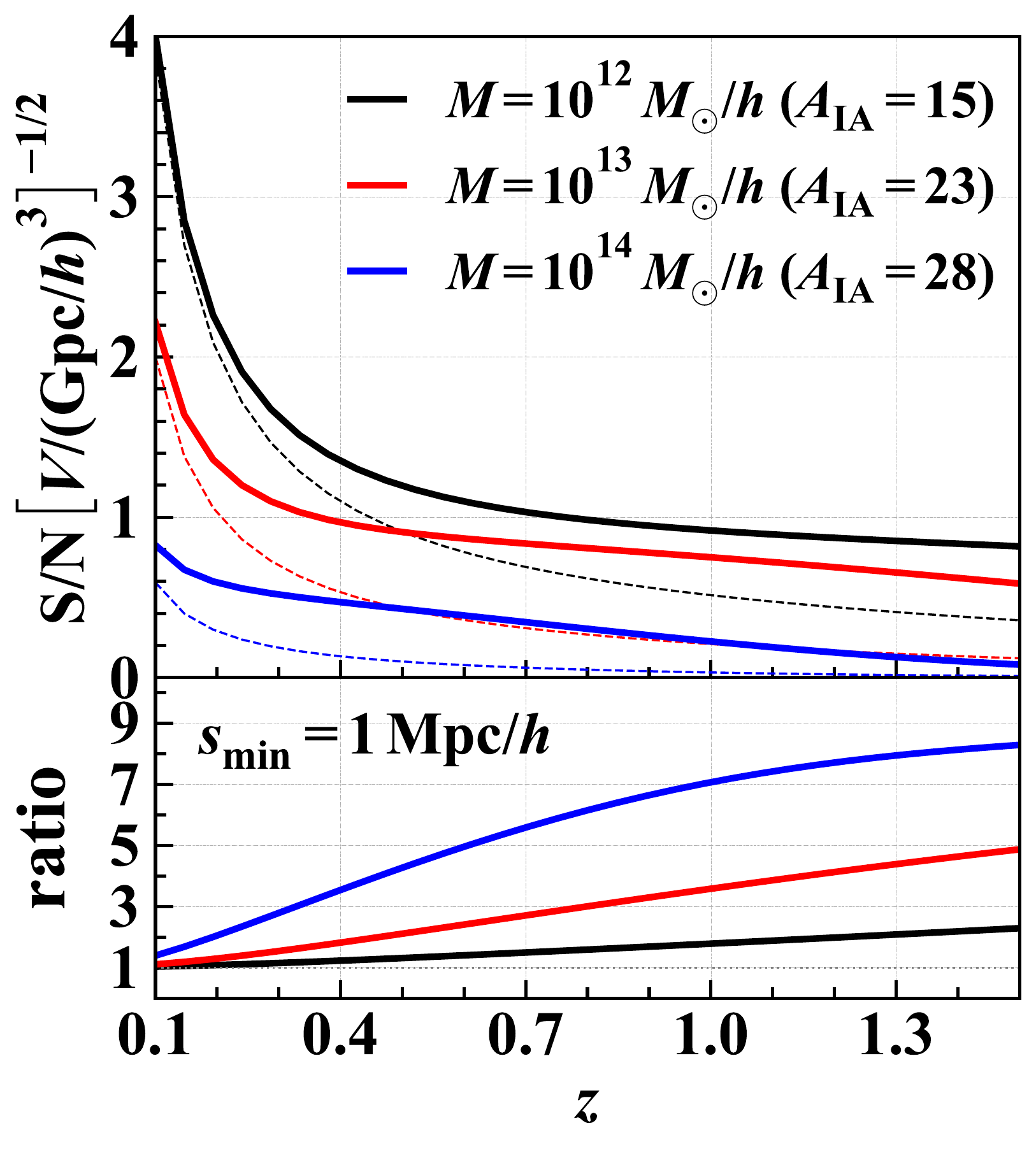}
\caption{
Same as Fig.~\ref{fig: SN smin} but we here show the results as a function of mean redshift $z$ with fixing $s_{\rm min} = 1\, {\rm Mpc}/h$.
The parameters used in these calculations are the same as Fig.~\ref{fig: SN smin}, except for $s_{\rm min}$.
}
\label{fig: SN zred}
\end{figure}

Let us next consider the redshift dependence of the signal-to-noise ratio, shown in Fig.~\ref{fig: SN zred}. Here, we fix the minimum separation to $s_{\rm min}=1\, {\rm Mpc}/h$.
Meanings of the line types are the same as shown in Fig.~\ref{fig: SN smin}. Since the signal-to-noise ratio shown here is normalized by the square root of the volume, the result monotonically decreases with the mean redshift.
As we have previously seen, the signal-to-noise ratio becomes smaller as increasing the halo mass, and this trend remains unchanged irrespective of the redshift.
Similar to Fig.~\ref{fig: SN smin}, we also compute the ratio of the results shown in solid to those in dotted lines, and plot them in the bottom panel.
Interestingly, we find that the gravitational redshift from the halo potential enhances the signal-to-noise ratio as increasing the redshift or halo mass.
This trend originates from the behaviour of the NFW profile, whose potential depth gets larger as increasing redshift and halo mass (see Appendix~E in \citet{2020MNRAS.498..981S}).

Finally, a naive impression from Figs.~\ref{fig: SN smin} and \ref{fig: SN zred} may be that even with an enhanced signal by the gravitational redshift effect from the halo potential, the detection of GI dipole is difficult or marginal in future surveys. Indeed, setting the the survey volume to $5\, ({\rm Gpc}/h)^{3}$ as a representative value of the upcoming surveys, the signal-to-noise ratio just reaches ${\rm S/N} \approx 2.5$ at $z\approx 0.5$ for $M=10^{12}\,M_{\odot}/h$. Nevertheless, a measurement of the GI dipole would be still valuable to detect the gravitational redshift effect. It was shown in \citet{2020ApJ...891L..42T} that the combination of GI correlation with the GG correlation significantly improves the measurement of baryon acoustic oscillations and redshift-space distortions, leading to tighter constraints on cosmological parameters \citep[see also][]{2022PhRvD.106d3523O}. Hence, we anticipate that combining the GI dipole with the GG cross dipole significantly improves the signal-to-noise ratio for gravitational redshift effect.
A great emphasis is that the GG dipole needs two galaxy samples with different bias parameters, whereas a non-zero GI dipole is obtained even with a single galaxy population. Thus, increasing a number of different galaxy populations to cross correlate, the improvement of the signal-to-noise ratio for GI dipole would be better than that for the GG cross dipole. In this respect, combining GG with GI dipole would be beneficial to enhance the detectability of the gravitational redshift effect. A quantitative estimate of the improvement for signal-to-noise ratios depends on the specification of each galaxy survey and the data analysis strategy to combine multiple galaxy populations. We will leave it to future work.

\section{Summary}
\label{sec: summary}

The observed galaxy distribution via spectroscopic surveys appears distorted due to not only the well-known contribution from the Doppler effect but also the contributions arising from the other special and general relativistic effects.
Such relativistic effects are known to produce the asymmetry in the galaxy-galaxy cross-correlation (GG correlation) between different biased objects.
Our recent works have shown that the gravitational redshift effect arising from the halo potential dominates the GG dipole at $s\lesssim 10\, {\rm Mpc}/h$~\citep{2019MNRAS.483.2671B,2020MNRAS.498..981S}, with the sign of its amplitude flipped.

In this paper, employing the perturbative approach as we have previously done in the GG dipole, we have constructed an analytical model for the cross-correlation function between the galaxy and intrinsic alignments (IAs) of galaxy shapes (GI correlation), taking major observational relativistic effects (i.e., Doppler and gravitational redshift effects) into account~\citep{2022MNRAS.511.2732S}.
Adopting the linear alignment model that linearly relates the ellipticity field to the tidal field of large-scale structure~\citep{2001MNRAS.320L...7C,2004PhRvD..70f3526H}, we have derived analytical expressions for the GI dipole correlation function~(see Eqs.~(\ref{eq: dipole real})--(\ref{eq: dipole eps})), where the wide-angle corrections, arising not only from the Doppler effect but also from the galaxy shape projected along the line of sight, are also incorporated in a proper way at the leading order beyond the plane-parallel limit. As a result, the GI dipole is shown to be non-zero even in the real space.

An important aspect of considering the GI dipole is that unlike the GG dipole that needs at least two different biased galaxy samples to detect the gravitational redshift effect from halo potential, even a single galaxy population leads to a non-zero signal. In this respect, the GI dipole potentially offers a good probe of the gravitational redshift effect, and thus can be used for an alternative test of gravity~\cite[e.g.,][for the recently proposed tests of gravity theory using the GG dipole]{2018JCAP...05..061B,2020JCAP...08..004B,2021arXiv211207727S}.

We found that the GI dipole at small scales ($s\lesssim 10\, {\rm Mpc}/h$) 
is dominated by the gravitational redshift effect from the halo potential. One notable point is that unlike the GG cross dipole, the sign flip of the amplitude of the GI dipole does not occur. The main reason for this comes from the fact that in contrast to the GG cross correlation, the relative positions between the ellipticity and density fields do not interchange in redshift space. This is because the galaxy ellipticity field is not affected by the Doppler and gravitational redshift effects at leading order.

Based on the analytical GI dipole, we have also estimated the signal-to-noise ratio and briefly discussed its future detectability. Extending the previous formula for the GG cross dipole, we have derived a new expression for the covariance matrix for the GI dipole, accounting properly for the azimuthal angle dependence of the GI correlation. Plugging this formula into the definition of the signal-to-noise ratio, we found that the signal-to-noise ratio for GI dipole increases as decreasing the minimum separation and/or halo mass. The improvement of the signal-to-noise ratio is also seen as decreasing the redshift, but a relative contribution of the gravitational redshift effect from the halo potential turns out to be suppressed (see Fig.~\ref{fig: SN zred}), indicating that a careful strategy to measure a non-zero GI dipole signal is needed.

In this paper, we considered the detectability of the gravitational redshift effect from the GI dipole alone. As we mentioned above, combined analysis with the GG correlations as well as the GI octupole could improve the detection of gravitational redshift effect, and help disentangling the degeneracy between model parameters. A more quantitative forecast study needs a specification of each galaxy survey as well as data analysis strategy on how to combine multiple galaxy surveys. In doing so, a proper account of the systematic effects that can potentially affect the detectability of the gravitational redshift effect is important, including the transverse Doppler effect and magnification bias~\citep{2011PhRvD..84f3505B,2017PhRvD..95d3530H}.
As another aspect, we may exploit a method to use GI dipole as a test of fundamental physics, and this would also be an interesting subject particularly through a precision measurement of the gravitational redshift effect~\citep[e.g.,][for the test of local position invariance by the GG dipole]{2021arXiv211207727S}.
Finally, toward a practical measurement of the GI dipole and a detection of the gravitational redshift effect, one has to consider a more realistic situation taking into account the observational systematics that have been ignored in this paper. A study with galaxy mock catalogues would be thus an important direction~\citep[e.g.,][]{2019MNRAS.483.2671B}. 
Comparing the results from mock catalogues with analytical predictions, the analytical model presented in this paper will be refined and calibrated, enabling us to practically measure and detect the non-zero dipole from the gravitational redshift effect in future observations. We will tackle these issues in our future work.

\section*{Acknowledgments}
We would like to thank Kazuyuki Akitsu for useful discussions.
We thank the Yukawa Institute for Theoretical Physics (YITP) at Kyoto University, where this work was completed during the YITP workshop YITP-T-21-06 on ``Galaxy shape statistics and cosmology''.
SS is supported by JSPS Overseas Research Fellowships.
TO acknowledges support from the Ministry of Science and Technology of Taiwan under Grant Nos. MOST 110-2112-M-001-045- and 111-2112-M-001-061-
and the Career Development Award, Academia Sinica (AS-CDA-108-M02) for the period of 2019 to 2023.
AT acknowledges the support from MEXT/JSPS KAKENHI Grant Nos. JP16H03977, JP17H06359, and JP20H05861.
AT was also supported by JST AIP Acceleration Research Grant No. JP20317829, Japan.

\section*{Data availability}
The data underlying this article are available in the article.

\bibliographystyle{mnras}
\bibliography{ref}

\begin{thebibliography}{}
\makeatletter
\relax
\def\mn@urlcharsother{\let\do\@makeother \do\$\do\&\do\#\do\^\do\_\do\%\do\~}
\def\mn@doi{\begingroup\mn@urlcharsother \@ifnextchar [ {\mn@doi@}
  {\mn@doi@[]}}
\def\mn@doi@[#1]#2{\def\@tempa{#1}\ifx\@tempa\@empty \href
  {http://dx.doi.org/#2} {doi:#2}\else \href {http://dx.doi.org/#2} {#1}\fi
  \endgroup}
\def\mn@eprint#1#2{\mn@eprint@#1:#2::\@nil}
\def\mn@eprint@arXiv#1{\href {http://arxiv.org/abs/#1} {{\tt arXiv:#1}}}
\def\mn@eprint@dblp#1{\href {http://dblp.uni-trier.de/rec/bibtex/#1.xml}
  {dblp:#1}}
\def\mn@eprint@#1:#2:#3:#4\@nil{\def\@tempa {#1}\def\@tempb {#2}\def\@tempc
  {#3}\ifx \@tempc \@empty \let \@tempc \@tempb \let \@tempb \@tempa \fi \ifx
  \@tempb \@empty \def\@tempb {arXiv}\fi \@ifundefined
  {mn@eprint@\@tempb}{\@tempb:\@tempc}{\expandafter \expandafter \csname
  mn@eprint@\@tempb\endcsname \expandafter{\@tempc}}}

\bibitem[\protect\citeauthoryear{{Bertacca}, {Maartens}, {Raccanelli}  \&
  {Clarkson}}{{Bertacca} et~al.}{2012}]{2012JCAP...10..025B}
{Bertacca} D.,  {Maartens} R.,  {Raccanelli} A.,   {Clarkson} C.,  2012,
  \mn@doi [\jcap] {10.1088/1475-7516/2012/10/025}, \href
  {https://ui.adsabs.harvard.edu/abs/2012JCAP...10..025B} {2012, 025}

\bibitem[\protect\citeauthoryear{{Beutler} \& {Di Dio}}{{Beutler} \& {Di
  Dio}}{2020}]{2020JCAP...07..048B}
{Beutler} F.,  {Di Dio} E.,  2020, \mn@doi [\jcap]
  {10.1088/1475-7516/2020/07/048}, \href
  {https://ui.adsabs.harvard.edu/abs/2020JCAP...07..048B} {2020, 048}

\bibitem[\protect\citeauthoryear{{Bonvin} \& {Durrer}}{{Bonvin} \&
  {Durrer}}{2011}]{2011PhRvD..84f3505B}
{Bonvin} C.,  {Durrer} R.,  2011, \mn@doi [\prd] {10.1103/PhysRevD.84.063505},
  \href {https://ui.adsabs.harvard.edu/abs/2011PhRvD..84f3505B} {84, 063505}

\bibitem[\protect\citeauthoryear{{Bonvin} \& {Fleury}}{{Bonvin} \&
  {Fleury}}{2018}]{2018JCAP...05..061B}
{Bonvin} C.,  {Fleury} P.,  2018, \mn@doi [\jcap]
  {10.1088/1475-7516/2018/05/061}, \href
  {https://ui.adsabs.harvard.edu/abs/2018JCAP...05..061B} {2018, 061}

\bibitem[\protect\citeauthoryear{{Bonvin}, {Hui}  \& {Gazta{\~n}aga}}{{Bonvin}
  et~al.}{2014}]{2014PhRvD..89h3535B}
{Bonvin} C.,  {Hui} L.,   {Gazta{\~n}aga} E.,  2014, \mn@doi [\prd]
  {10.1103/PhysRevD.89.083535}, \href
  {https://ui.adsabs.harvard.edu/abs/2014PhRvD..89h3535B} {89, 083535}

\bibitem[\protect\citeauthoryear{{Bonvin}, {Hui}  \& {Gaztanaga}}{{Bonvin}
  et~al.}{2016}]{2016JCAP...08..021B}
{Bonvin} C.,  {Hui} L.,   {Gaztanaga} E.,  2016, \mn@doi [\jcap]
  {10.1088/1475-7516/2016/08/021}, \href
  {https://ui.adsabs.harvard.edu/abs/2016JCAP...08..021B} {2016, 021}

\bibitem[\protect\citeauthoryear{{Bonvin}, {Oliveira Franco}  \&
  {Fleury}}{{Bonvin} et~al.}{2020}]{2020JCAP...08..004B}
{Bonvin} C.,  {Oliveira Franco} F.,   {Fleury} P.,  2020, \mn@doi [\jcap]
  {10.1088/1475-7516/2020/08/004}, \href
  {https://ui.adsabs.harvard.edu/abs/2020JCAP...08..004B} {2020, 004}

\bibitem[\protect\citeauthoryear{{Borzyszkowski}, {Bertacca}  \&
  {Porciani}}{{Borzyszkowski} et~al.}{2017}]{2017MNRAS.471.3899B}
{Borzyszkowski} M.,  {Bertacca} D.,   {Porciani} C.,  2017, \mn@doi [\mnras]
  {10.1093/mnras/stx1423}, \href
  {https://ui.adsabs.harvard.edu/abs/2017MNRAS.471.3899B} {471, 3899}

\bibitem[\protect\citeauthoryear{{Breton}, {Rasera}, {Taruya}, {Lacombe}  \&
  {Saga}}{{Breton} et~al.}{2019}]{2019MNRAS.483.2671B}
{Breton} M.-A.,  {Rasera} Y.,  {Taruya} A.,  {Lacombe} O.,   {Saga} S.,  2019,
  \mn@doi [\mnras] {10.1093/mnras/sty3206}, \href
  {https://ui.adsabs.harvard.edu/\#abs/2019MNRAS.483.2671B} {483, 2671}

\bibitem[\protect\citeauthoryear{{Catelan} \& {Porciani}}{{Catelan} \&
  {Porciani}}{2001}]{2001MNRAS.323..713C}
{Catelan} P.,  {Porciani} C.,  2001, \mn@doi [\mnras]
  {10.1046/j.1365-8711.2001.04250.x}, \href
  {https://ui.adsabs.harvard.edu/abs/2001MNRAS.323..713C} {323, 713}

\bibitem[\protect\citeauthoryear{{Catelan}, {Kamionkowski}  \&
  {Blandford}}{{Catelan} et~al.}{2001}]{2001MNRAS.320L...7C}
{Catelan} P.,  {Kamionkowski} M.,   {Blandford} R.~D.,  2001, \mn@doi [\mnras]
  {10.1046/j.1365-8711.2001.04105.x}, \href
  {https://ui.adsabs.harvard.edu/abs/2001MNRAS.320L...7C} {320, L7}

\bibitem[\protect\citeauthoryear{{Challinor} \& {Lewis}}{{Challinor} \&
  {Lewis}}{2011}]{2011PhRvD..84d3516C}
{Challinor} A.,  {Lewis} A.,  2011, \mn@doi [\prd]
  {10.1103/PhysRevD.84.043516}, \href
  {https://ui.adsabs.harvard.edu/abs/2011PhRvD..84d3516C} {84, 043516}

\bibitem[\protect\citeauthoryear{{Chisari} \& {Dvorkin}}{{Chisari} \&
  {Dvorkin}}{2013}]{2013JCAP...12..029C}
{Chisari} N.~E.,  {Dvorkin} C.,  2013, \mn@doi [\jcap]
  {10.1088/1475-7516/2013/12/029}, \href
  {https://ui.adsabs.harvard.edu/abs/2013JCAP...12..029C} {2013, 029}

\bibitem[\protect\citeauthoryear{{Chuang}, {Okumura}  \& {Shirasaki}}{{Chuang}
  et~al.}{2022}]{2022MNRAS.515.4464C}
{Chuang} Y.-T.,  {Okumura} T.,   {Shirasaki} M.,  2022, \mn@doi [\mnras]
  {10.1093/mnras/stac2029}, \href
  {https://ui.adsabs.harvard.edu/abs/2022MNRAS.515.4464C} {515, 4464}

\bibitem[\protect\citeauthoryear{{Clarkson}, {de Weerd}, {Jolicoeur},
  {Maartens}  \& {Umeh}}{{Clarkson} et~al.}{2019}]{2019MNRAS.486L.101C}
{Clarkson} C.,  {de Weerd} E.~M.,  {Jolicoeur} S.,  {Maartens} R.,   {Umeh} O.,
   2019, \mn@doi [\mnras] {10.1093/mnrasl/slz066}, \href
  {https://ui.adsabs.harvard.edu/abs/2019MNRAS.486L.101C} {486, L101}

\bibitem[\protect\citeauthoryear{{Coates}, {Adamek}, {Bull}, {Guandalin}  \&
  {Clarkson}}{{Coates} et~al.}{2021}]{2021MNRAS.504.3534C}
{Coates} L.,  {Adamek} J.,  {Bull} P.,  {Guandalin} C.,   {Clarkson} C.,  2021,
  \mn@doi [\mnras] {10.1093/mnras/stab1076}, \href
  {https://ui.adsabs.harvard.edu/abs/2021MNRAS.504.3534C} {504, 3534}

\bibitem[\protect\citeauthoryear{{Crittenden}, {Natarajan}, {Pen}  \&
  {Theuns}}{{Crittenden} et~al.}{2001}]{2001ApJ...559..552C}
{Crittenden} R.~G.,  {Natarajan} P.,  {Pen} U.-L.,   {Theuns} T.,  2001,
  \mn@doi [\apj] {10.1086/322370}, \href
  {https://ui.adsabs.harvard.edu/abs/2001ApJ...559..552C} {559, 552}

\bibitem[\protect\citeauthoryear{{Croft}}{{Croft}}{2013}]{2013MNRAS.434.3008C}
{Croft} R. A.~C.,  2013, \mn@doi [\mnras] {10.1093/mnras/stt1223}, \href
  {https://ui.adsabs.harvard.edu/abs/2013MNRAS.434.3008C} {434, 3008}

\bibitem[\protect\citeauthoryear{{Croft} \& {Metzler}}{{Croft} \&
  {Metzler}}{2000}]{2000ApJ...545..561C}
{Croft} R. A.~C.,  {Metzler} C.~A.,  2000, \mn@doi [\apj] {10.1086/317856},
  \href {https://ui.adsabs.harvard.edu/abs/2000ApJ...545..561C} {545, 561}

\bibitem[\protect\citeauthoryear{{Di Dio} \& {Seljak}}{{Di Dio} \&
  {Seljak}}{2019}]{2019JCAP...04..050D}
{Di Dio} E.,  {Seljak} U.,  2019, \mn@doi [\jcap]
  {10.1088/1475-7516/2019/04/050}, \href
  {https://ui.adsabs.harvard.edu/abs/2019JCAP...04..050D} {2019, 050}

\bibitem[\protect\citeauthoryear{{Gaztanaga}, {Bonvin}  \& {Hui}}{{Gaztanaga}
  et~al.}{2017}]{2017JCAP...01..032G}
{Gaztanaga} E.,  {Bonvin} C.,   {Hui} L.,  2017, \mn@doi [\jcap]
  {10.1088/1475-7516/2017/01/032}, \href
  {https://ui.adsabs.harvard.edu/abs/2017JCAP...01..032G} {2017, 032}

\bibitem[\protect\citeauthoryear{{Guandalin}, {Adamek}, {Bull}, {Clarkson},
  {Abramo}  \& {Coates}}{{Guandalin} et~al.}{2021}]{2021MNRAS.501.2547G}
{Guandalin} C.,  {Adamek} J.,  {Bull} P.,  {Clarkson} C.,  {Abramo} L.~R.,
  {Coates} L.,  2021, \mn@doi [\mnras] {10.1093/mnras/staa3890}, \href
  {https://ui.adsabs.harvard.edu/abs/2021MNRAS.501.2547G} {501, 2547}

\bibitem[\protect\citeauthoryear{{Hall} \& {Bonvin}}{{Hall} \&
  {Bonvin}}{2017}]{2017PhRvD..95d3530H}
{Hall} A.,  {Bonvin} C.,  2017, \mn@doi [\prd] {10.1103/PhysRevD.95.043530},
  \href {https://ui.adsabs.harvard.edu/abs/2017PhRvD..95d3530H} {95, 043530}

\bibitem[\protect\citeauthoryear{{Hamilton}}{{Hamilton}}{1992}]{1992ApJ...385L...5H}
{Hamilton} A.~J.~S.,  1992, \mn@doi [\apjl] {10.1086/186264}, \href
  {https://ui.adsabs.harvard.edu/abs/1992ApJ...385L...5H} {385, L5}

\bibitem[\protect\citeauthoryear{{Hamilton} \& {Culhane}}{{Hamilton} \&
  {Culhane}}{1996}]{1996MNRAS.278...73H}
{Hamilton} A.~J.~S.,  {Culhane} M.,  1996, \mn@doi [\mnras]
  {10.1093/mnras/278.1.73}, \href
  {https://ui.adsabs.harvard.edu/abs/1996MNRAS.278...73H} {278, 73}

\bibitem[\protect\citeauthoryear{{Heavens}, {Refregier}  \&
  {Heymans}}{{Heavens} et~al.}{2000}]{2000MNRAS.319..649H}
{Heavens} A.,  {Refregier} A.,   {Heymans} C.,  2000, \mn@doi [\mnras]
  {10.1046/j.1365-8711.2000.03907.x}, \href
  {https://ui.adsabs.harvard.edu/abs/2000MNRAS.319..649H} {319, 649}

\bibitem[\protect\citeauthoryear{{Hikage}, {Mandelbaum}, {Takada}  \&
  {Spergel}}{{Hikage} et~al.}{2013}]{2013MNRAS.435.2345H}
{Hikage} C.,  {Mandelbaum} R.,  {Takada} M.,   {Spergel} D.~N.,  2013, \mn@doi
  [\mnras] {10.1093/mnras/stt1446}, \href
  {https://ui.adsabs.harvard.edu/abs/2013MNRAS.435.2345H} {435, 2345}

\bibitem[\protect\citeauthoryear{{Hirata} \& {Seljak}}{{Hirata} \&
  {Seljak}}{2004}]{2004PhRvD..70f3526H}
{Hirata} C.~M.,  {Seljak} U.,  2004, \mn@doi [\prd]
  {10.1103/PhysRevD.70.063526}, \href
  {https://ui.adsabs.harvard.edu/abs/2004PhRvD..70f3526H} {70, 063526}

\bibitem[\protect\citeauthoryear{{Hirata}, {Mandelbaum}, {Ishak}, {Seljak},
  {Nichol}, {Pimbblet}, {Ross}  \& {Wake}}{{Hirata}
  et~al.}{2007}]{2007MNRAS.381.1197H}
{Hirata} C.~M.,  {Mandelbaum} R.,  {Ishak} M.,  {Seljak} U.,  {Nichol} R.,
  {Pimbblet} K.~A.,  {Ross} N.~P.,   {Wake} D.,  2007, \mn@doi [\mnras]
  {10.1111/j.1365-2966.2007.12312.x}, \href
  {https://ui.adsabs.harvard.edu/abs/2007MNRAS.381.1197H} {381, 1197}

\bibitem[\protect\citeauthoryear{{Hui}, {Gazta{\~n}aga}  \& {Loverde}}{{Hui}
  et~al.}{2008}]{2008PhRvD..77f3526H}
{Hui} L.,  {Gazta{\~n}aga} E.,   {Loverde} M.,  2008, \mn@doi [\prd]
  {10.1103/PhysRevD.77.063526}, \href
  {https://ui.adsabs.harvard.edu/abs/2008PhRvD..77f3526H} {77, 063526}

\bibitem[\protect\citeauthoryear{{Joachimi}, {Mandelbaum}, {Abdalla}  \&
  {Bridle}}{{Joachimi} et~al.}{2011}]{2011A&A...527A..26J}
{Joachimi} B.,  {Mandelbaum} R.,  {Abdalla} F.~B.,   {Bridle} S.~L.,  2011,
  \mn@doi [\aap] {10.1051/0004-6361/201015621}, \href
  {https://ui.adsabs.harvard.edu/abs/2011A&A...527A..26J} {527, A26}

\bibitem[\protect\citeauthoryear{{Johnston} et~al.,}{{Johnston}
  et~al.}{2019}]{2019A&A...624A..30J}
{Johnston} H.,  et~al., 2019, \mn@doi [\aap] {10.1051/0004-6361/201834714},
  \href {https://ui.adsabs.harvard.edu/abs/2019A&A...624A..30J} {624, A30}

\bibitem[\protect\citeauthoryear{{Kaiser}}{{Kaiser}}{1987}]{1987MNRAS.227....1K}
{Kaiser} N.,  1987, \mn@doi [\mnras] {10.1093/mnras/227.1.1}, \href
  {https://ui.adsabs.harvard.edu/abs/1987MNRAS.227....1K} {227, 1}

\bibitem[\protect\citeauthoryear{{Komatsu} et~al.,}{{Komatsu}
  et~al.}{2011}]{2011ApJS..192...18K}
{Komatsu} E.,  et~al., 2011, \mn@doi [¥apjs] {10.1088/0067-0049/192/2/18},
  \href {https://ui.adsabs.harvard.edu/abs/2011ApJS..192...18K} {192, 18}

\bibitem[\protect\citeauthoryear{{Kurita}, {Takada}, {Nishimichi}, {Takahashi},
  {Osato}  \& {Kobayashi}}{{Kurita} et~al.}{2021}]{2021MNRAS.501..833K}
{Kurita} T.,  {Takada} M.,  {Nishimichi} T.,  {Takahashi} R.,  {Osato} K.,
  {Kobayashi} Y.,  2021, \mn@doi [\mnras] {10.1093/mnras/staa3625}, \href
  {https://ui.adsabs.harvard.edu/abs/2021MNRAS.501..833K} {501, 833}

\bibitem[\protect\citeauthoryear{{Lepori}, {Di Dio}, {Villa}  \&
  {Viel}}{{Lepori} et~al.}{2018}]{2018JCAP...05..043L}
{Lepori} F.,  {Di Dio} E.,  {Villa} E.,   {Viel} M.,  2018, \mn@doi [\jcap]
  {10.1088/1475-7516/2018/05/043}, \href
  {https://ui.adsabs.harvard.edu/abs/2018JCAP...05..043L} {2018, 043}

\bibitem[\protect\citeauthoryear{{Maartens}, {Jolicoeur}, {Umeh}, {De Weerd},
  {Clarkson}  \& {Camera}}{{Maartens} et~al.}{2020}]{2020JCAP...03..065M}
{Maartens} R.,  {Jolicoeur} S.,  {Umeh} O.,  {De Weerd} E.~M.,  {Clarkson} C.,
   {Camera} S.,  2020, \mn@doi [\jcap] {10.1088/1475-7516/2020/03/065}, \href
  {https://ui.adsabs.harvard.edu/abs/2020JCAP...03..065M} {2020, 065}

\bibitem[\protect\citeauthoryear{{Mandelbaum}, {Hirata}, {Ishak}, {Seljak}  \&
  {Brinkmann}}{{Mandelbaum} et~al.}{2006}]{2006MNRAS.367..611M}
{Mandelbaum} R.,  {Hirata} C.~M.,  {Ishak} M.,  {Seljak} U.,   {Brinkmann} J.,
  2006, \mn@doi [\mnras] {10.1111/j.1365-2966.2005.09946.x}, \href
  {https://ui.adsabs.harvard.edu/abs/2006MNRAS.367..611M} {367, 611}

\bibitem[\protect\citeauthoryear{{Matsubara}}{{Matsubara}}{2000a}]{2000ApJ...535....1M}
{Matsubara} T.,  2000a, \mn@doi [\apj] {10.1086/308827}, \href
  {https://ui.adsabs.harvard.edu/abs/2000ApJ...535....1M} {535, 1}

\bibitem[\protect\citeauthoryear{{Matsubara}}{{Matsubara}}{2000b}]{2000ApJ...537L..77M}
{Matsubara} T.,  2000b, \mn@doi [\apjl] {10.1086/312762}, \href
  {https://ui.adsabs.harvard.edu/abs/2000ApJ...537L..77M} {537, L77}

\bibitem[\protect\citeauthoryear{{McDonald}}{{McDonald}}{2009}]{2009JCAP...11..026M}
{McDonald} P.,  2009, \mn@doi [\jcap] {10.1088/1475-7516/2009/11/026}, \href
  {https://ui.adsabs.harvard.edu/\#abs/2009JCAP...11..026M} {2009, 026}

\bibitem[\protect\citeauthoryear{{Navarro}, {Frenk}  \& {White}}{{Navarro}
  et~al.}{1996}]{1996ApJ...462..563N}
{Navarro} J.~F.,  {Frenk} C.~S.,   {White} S. D.~M.,  1996, \mn@doi [\apj]
  {10.1086/177173}, \href
  {https://ui.adsabs.harvard.edu/abs/1996ApJ...462..563N} {462, 563}

\bibitem[\protect\citeauthoryear{{Novikov}}{{Novikov}}{1969}]{1969JETP...30..512N}
{Novikov} E.~A.,  1969, Soviet Journal of Experimental and Theoretical Physics,
  \href {https://ui.adsabs.harvard.edu/abs/1969JETP...30..512N} {30, 512}

\bibitem[\protect\citeauthoryear{{Okumura} \& {Jing}}{{Okumura} \&
  {Jing}}{2009}]{2009ApJ...694L..83O}
{Okumura} T.,  {Jing} Y.~P.,  2009, \mn@doi [\apjl]
  {10.1088/0004-637X/694/1/L83}, \href
  {https://ui.adsabs.harvard.edu/abs/2009ApJ...694L..83O} {694, L83}

\bibitem[\protect\citeauthoryear{{Okumura} \& {Taruya}}{{Okumura} \&
  {Taruya}}{2020}]{2020MNRAS.493L.124O}
{Okumura} T.,  {Taruya} A.,  2020, \mn@doi [\mnras] {10.1093/mnrasl/slaa024},
  \href {https://ui.adsabs.harvard.edu/abs/2020MNRAS.493L.124O} {493, L124}

\bibitem[\protect\citeauthoryear{{Okumura} \& {Taruya}}{{Okumura} \&
  {Taruya}}{2022}]{2022PhRvD.106d3523O}
{Okumura} T.,  {Taruya} A.,  2022, \mn@doi [\prd]
  {10.1103/PhysRevD.106.043523}, \href
  {https://ui.adsabs.harvard.edu/abs/2022PhRvD.106d3523O} {106, 043523}

\bibitem[\protect\citeauthoryear{{Okumura}, {Jing}  \& {Li}}{{Okumura}
  et~al.}{2009}]{2009ApJ...694..214O}
{Okumura} T.,  {Jing} Y.~P.,   {Li} C.,  2009, \mn@doi [\apj]
  {10.1088/0004-637X/694/1/214}, \href
  {https://ui.adsabs.harvard.edu/abs/2009ApJ...694..214O} {694, 214}

\bibitem[\protect\citeauthoryear{{Okumura}, {Taruya}  \&
  {Nishimichi}}{{Okumura} et~al.}{2019}]{2019PhRvD.100j3507O}
{Okumura} T.,  {Taruya} A.,   {Nishimichi} T.,  2019, \mn@doi [\prd]
  {10.1103/PhysRevD.100.103507}, \href
  {https://ui.adsabs.harvard.edu/abs/2019PhRvD.100j3507O} {100, 103507}

\bibitem[\protect\citeauthoryear{{Okumura}, {Taruya}  \&
  {Nishimichi}}{{Okumura} et~al.}{2020}]{2020MNRAS.494..694O}
{Okumura} T.,  {Taruya} A.,   {Nishimichi} T.,  2020, \mn@doi [\mnras]
  {10.1093/mnras/staa718}, \href
  {https://ui.adsabs.harvard.edu/abs/2020MNRAS.494..694O} {494, 694}

\bibitem[\protect\citeauthoryear{{Pyne} \& {Birkinshaw}}{{Pyne} \&
  {Birkinshaw}}{2004}]{2004MNRAS.348..581P}
{Pyne} T.,  {Birkinshaw} M.,  2004, \mn@doi [\mnras]
  {10.1111/j.1365-2966.2004.07362.x}, \href
  {https://ui.adsabs.harvard.edu/abs/2004MNRAS.348..581P} {348, 581}

\bibitem[\protect\citeauthoryear{{Reimberg}, {Bernardeau}  \&
  {Pitrou}}{{Reimberg} et~al.}{2016}]{2016JCAP...01..048R}
{Reimberg} P.,  {Bernardeau} F.,   {Pitrou} C.,  2016, \mn@doi [\jcap]
  {10.1088/1475-7516/2016/01/048}, \href
  {https://ui.adsabs.harvard.edu/abs/2016JCAP...01..048R} {2016, 048}

\bibitem[\protect\citeauthoryear{{Saga}, {Taruya}, {Breton}  \&
  {Rasera}}{{Saga} et~al.}{2020}]{2020MNRAS.498..981S}
{Saga} S.,  {Taruya} A.,  {Breton} M.-A.,   {Rasera} Y.,  2020, \mn@doi
  [\mnras] {10.1093/mnras/staa2232}, \href
  {https://ui.adsabs.harvard.edu/abs/2020MNRAS.498..981S} {498, 981}

\bibitem[\protect\citeauthoryear{{Saga}, {Taruya}, {Breton}  \&
  {Rasera}}{{Saga} et~al.}{2021}]{2021arXiv211207727S}
{Saga} S.,  {Taruya} A.,  {Breton} M.-A.,   {Rasera} Y.,  2021, arXiv e-prints,
  \href {https://ui.adsabs.harvard.edu/abs/2021arXiv211207727S} {p.
  arXiv:2112.07727}

\bibitem[\protect\citeauthoryear{{Saga}, {Taruya}, {Rasera}  \&
  {Breton}}{{Saga} et~al.}{2022}]{2022MNRAS.511.2732S}
{Saga} S.,  {Taruya} A.,  {Rasera} Y.,   {Breton} M.-A.,  2022, \mn@doi
  [\mnras] {10.1093/mnras/stac186}, \href
  {https://ui.adsabs.harvard.edu/abs/2022MNRAS.511.2732S} {511, 2732}

\bibitem[\protect\citeauthoryear{Salem \& Wio}{Salem \& Wio}{1989}]{Salem_1989}
Salem L.~D.,  Wio H.~S.,  1989, \mn@doi [Journal of Physics A: Mathematical and
  General] {10.1088/0305-4470/22/20/012}, 22, 4331

\bibitem[\protect\citeauthoryear{{Samuroff} et~al.,}{{Samuroff}
  et~al.}{2019}]{2019MNRAS.489.5453S}
{Samuroff} S.,  et~al., 2019, \mn@doi [\mnras] {10.1093/mnras/stz2197}, \href
  {https://ui.adsabs.harvard.edu/abs/2019MNRAS.489.5453S} {489, 5453}

\bibitem[\protect\citeauthoryear{{Sasaki}}{{Sasaki}}{1987}]{1987MNRAS.228..653S}
{Sasaki} M.,  1987, \mn@doi [\mnras] {10.1093/mnras/228.3.653}, \href
  {https://ui.adsabs.harvard.edu/abs/1987MNRAS.228..653S} {228, 653}

\bibitem[\protect\citeauthoryear{{Schmidt} \& {Jeong}}{{Schmidt} \&
  {Jeong}}{2012}]{2012PhRvD..86h3513S}
{Schmidt} F.,  {Jeong} D.,  2012, \mn@doi [\prd] {10.1103/PhysRevD.86.083513},
  \href {https://ui.adsabs.harvard.edu/abs/2012PhRvD..86h3513S} {86, 083513}

\bibitem[\protect\citeauthoryear{{Schmitz}, {Hirata}, {Blazek}  \&
  {Krause}}{{Schmitz} et~al.}{2018}]{2018JCAP...07..030S}
{Schmitz} D.~M.,  {Hirata} C.~M.,  {Blazek} J.,   {Krause} E.,  2018, \mn@doi
  [\jcap] {10.1088/1475-7516/2018/07/030}, \href
  {https://ui.adsabs.harvard.edu/abs/2018JCAP...07..030S} {2018, 030}

\bibitem[\protect\citeauthoryear{{Shandarin} \& {Zeldovich}}{{Shandarin} \&
  {Zeldovich}}{1989}]{1989RvMP...61..185S}
{Shandarin} S.~F.,  {Zeldovich} Y.~B.,  1989, \mn@doi [Reviews of Modern
  Physics] {10.1103/RevModPhys.61.185}, \href
  {https://ui.adsabs.harvard.edu/abs/1989RvMP...61..185S} {61, 185}

\bibitem[\protect\citeauthoryear{{Sheth} \& {Tormen}}{{Sheth} \&
  {Tormen}}{1999}]{1999MNRAS.308..119S}
{Sheth} R.~K.,  {Tormen} G.,  1999, \mn@doi [\mnras]
  {10.1046/j.1365-8711.1999.02692.x}, \href
  {https://ui.adsabs.harvard.edu/abs/1999MNRAS.308..119S} {308, 119}

\bibitem[\protect\citeauthoryear{{Shiraishi}, {Taruya}, {Okumura}  \&
  {Akitsu}}{{Shiraishi} et~al.}{2021}]{2021MNRAS.503L...6S}
{Shiraishi} M.,  {Taruya} A.,  {Okumura} T.,   {Akitsu} K.,  2021, \mn@doi
  [\mnras] {10.1093/mnrasl/slab009}, \href
  {https://ui.adsabs.harvard.edu/abs/2021MNRAS.503L...6S} {503, L6}

\bibitem[\protect\citeauthoryear{{Singh}, {Mandelbaum}  \& {More}}{{Singh}
  et~al.}{2015}]{2015MNRAS.450.2195S}
{Singh} S.,  {Mandelbaum} R.,   {More} S.,  2015, \mn@doi [\mnras]
  {10.1093/mnras/stv778}, \href
  {https://ui.adsabs.harvard.edu/abs/2015MNRAS.450.2195S} {450, 2195}

\bibitem[\protect\citeauthoryear{{Szalay}, {Matsubara}  \& {Landy}}{{Szalay}
  et~al.}{1998}]{1998ApJ...498L...1S}
{Szalay} A.~S.,  {Matsubara} T.,   {Landy} S.~D.,  1998, \mn@doi [\apjl]
  {10.1086/311293}, \href
  {https://ui.adsabs.harvard.edu/abs/1998ApJ...498L...1S} {498, L1}

\bibitem[\protect\citeauthoryear{{Szapudi}}{{Szapudi}}{2004}]{2004ApJ...614...51S}
{Szapudi} I.,  2004, \mn@doi [\apj] {10.1086/423168}, \href
  {https://ui.adsabs.harvard.edu/abs/2004ApJ...614...51S} {614, 51}

\bibitem[\protect\citeauthoryear{{Tansella}, {Bonvin}, {Durrer}, {Ghosh}  \&
  {Sellentin}}{{Tansella} et~al.}{2018}]{2018JCAP...03..019T}
{Tansella} V.,  {Bonvin} C.,  {Durrer} R.,  {Ghosh} B.,   {Sellentin} E.,
  2018, \mn@doi [\jcap] {10.1088/1475-7516/2018/03/019}, \href
  {https://ui.adsabs.harvard.edu/abs/2018JCAP...03..019T} {2018, 019}

\bibitem[\protect\citeauthoryear{{Taruya} \& {Okumura}}{{Taruya} \&
  {Okumura}}{2020}]{2020ApJ...891L..42T}
{Taruya} A.,  {Okumura} T.,  2020, \mn@doi [\apjl] {10.3847/2041-8213/ab7934},
  \href {https://ui.adsabs.harvard.edu/abs/2020ApJ...891L..42T} {891, L42}

\bibitem[\protect\citeauthoryear{{Taruya}, {Saga}, {Breton}, {Rasera}  \&
  {Fujita}}{{Taruya} et~al.}{2020}]{2020MNRAS.491.4162T}
{Taruya} A.,  {Saga} S.,  {Breton} M.-A.,  {Rasera} Y.,   {Fujita} T.,  2020,
  \mn@doi [\mnras] {10.1093/mnras/stz3272}, \href
  {https://ui.adsabs.harvard.edu/abs/2020MNRAS.491.4162T} {491, 4162}

\bibitem[\protect\citeauthoryear{{Tonegawa} \& {Okumura}}{{Tonegawa} \&
  {Okumura}}{2022}]{2022ApJ...924L...3T}
{Tonegawa} M.,  {Okumura} T.,  2022, \mn@doi [\apjl]
  {10.3847/2041-8213/ac4246}, \href
  {https://ui.adsabs.harvard.edu/abs/2022ApJ...924L...3T} {924, L3}

\bibitem[\protect\citeauthoryear{{Troxel} \& {Ishak}}{{Troxel} \&
  {Ishak}}{2015}]{2015PhR...558....1T}
{Troxel} M.~A.,  {Ishak} M.,  2015, \mn@doi [\physrep]
  {10.1016/j.physrep.2014.11.001}, \href
  {https://ui.adsabs.harvard.edu/abs/2015PhR...558....1T} {558, 1}

\bibitem[\protect\citeauthoryear{{Yan}, {Raza}, {Van Waerbeke}, {Mead},
  {McCarthy}, {Tr{\"o}ster}  \& {Hinshaw}}{{Yan}
  et~al.}{2020}]{2020MNRAS.493.1120Y}
{Yan} Z.,  {Raza} N.,  {Van Waerbeke} L.,  {Mead} A.~J.,  {McCarthy} I.~G.,
  {Tr{\"o}ster} T.,   {Hinshaw} G.,  2020, \mn@doi [\mnras]
  {10.1093/mnras/staa295}, \href
  {https://ui.adsabs.harvard.edu/abs/2020MNRAS.493.1120Y} {493, 1120}

\bibitem[\protect\citeauthoryear{{Yao}, {Shan}, {Zhang}, {Kneib}  \&
  {Jullo}}{{Yao} et~al.}{2020}]{2020ApJ...904..135Y}
{Yao} J.,  {Shan} H.,  {Zhang} P.,  {Kneib} J.-P.,   {Jullo} E.,  2020, \mn@doi
  [\apj] {10.3847/1538-4357/abc175}, \href
  {https://ui.adsabs.harvard.edu/abs/2020ApJ...904..135Y} {904, 135}

\bibitem[\protect\citeauthoryear{{Yoo}}{{Yoo}}{2010}]{2010PhRvD..82h3508Y}
{Yoo} J.,  2010, \mn@doi [\prd] {10.1103/PhysRevD.82.083508}, \href
  {https://ui.adsabs.harvard.edu/abs/2010PhRvD..82h3508Y} {82, 083508}

\bibitem[\protect\citeauthoryear{{Yoo}}{{Yoo}}{2014}]{2014CQGra..31w4001Y}
{Yoo} J.,  2014, \mn@doi [Classical and Quantum Gravity]
  {10.1088/0264-9381/31/23/234001}, \href
  {https://ui.adsabs.harvard.edu/abs/2014CQGra..31w4001Y} {31, 234001}

\bibitem[\protect\citeauthoryear{{Yoo}, {Fitzpatrick}  \& {Zaldarriaga}}{{Yoo}
  et~al.}{2009}]{2009PhRvD..80h3514Y}
{Yoo} J.,  {Fitzpatrick} A.~L.,   {Zaldarriaga} M.,  2009, \mn@doi [\prd]
  {10.1103/PhysRevD.80.083514}, \href
  {https://ui.adsabs.harvard.edu/abs/2009PhRvD..80h3514Y} {80, 083514}

\bibitem[\protect\citeauthoryear{{Yoo}, {Hamaus}, {Seljak}  \&
  {Zaldarriaga}}{{Yoo} et~al.}{2012}]{2012arXiv1206.5809Y}
{Yoo} J.,  {Hamaus} N.,  {Seljak} U.,   {Zaldarriaga} M.,  2012, arXiv
  e-prints, \href {https://ui.adsabs.harvard.edu/abs/2012arXiv1206.5809Y} {p.
  arXiv:1206.5809}

\bibitem[\protect\citeauthoryear{{Zel'dovich}}{{Zel'dovich}}{1970}]{1970A&A.....5...84Z}
{Zel'dovich} Y.~B.,  1970, \aap, \href
  {https://ui.adsabs.harvard.edu/abs/1970A%26A.....5...84Z} {5, 84}

\makeatother
\end{thebibliography}

\appendix

\section{Summary of analytical expressions for GI correlation function}
\label{app: coefficients}

In this appendix, we summarize the coefficients related to the correlation function given in Sec.~\ref{sec: GI corr}, as well as the multipole moments of the correlation function.

In Sec.~\ref{app: full wa} and Sec.~\ref{app: multipoles}, we present the analytical expressions for the coefficients in Eq.~(\ref{eq: xi GI}) and the non-vanishing multipoles, respectively.
Since we use the spherical harmonic expansion in deriving the covariance matrix for the dipole of GI correlation, we also summarize the spherical harmonic coefficients in Sec.~\ref{app: coefficients Ylm}.

\subsection{Coefficients in GI correlation, $\alpha_\ell^{\rm(n)}$}
\label{app: full wa}

In Sec.~\ref{sec: GI corr}, introducing specifically the coordinate system in Fig.~\ref{fig: config}, the GI correlation function is given 
in the form of Eq.~(\ref{eq: xi GI}), with the dimensionless coefficient $\alpha^{(n)}_{\ell}$ characterizing the geometric dependence of the correlation function. This expression can be derived by substituting Eqs.~(\ref{eq: delta std + delta epsNL}) and (\ref{eq: gamma field}) into Eq.~(\ref{eq: def GI corr}). While the result apparently involves the three-dimensional Fourier integral, the angular integration can be partly performed, and with a help of the formulas summarized in Appendix~A in \citet{2022MNRAS.511.2732S}, the final expression is reduced to the one-dimensional integral form, as given in Eq.~(\ref{eq: xi GI}). The derivation is almost parallel to that of the GG correlation function in \citet{2022MNRAS.511.2732S}, and we below present only the resultant expressions of the coefficient $\alpha_\ell^{\rm (n)}$. In doing so, it is convenient to write the coefficients in a separate form as follows: 
\begin{align}
\alpha^{(n)}_{\ell} = b_{\rm K}\left[ b\, a^{(n)}_{\ell} + f\, b^{(n)}_{\ell} + \mathcal{M}s\, c^{(n)}_{\ell} - \frac{\phi_{\rm halo}}{aHs} \left( d^{(n)}_{\ell} + f\,e^{(n)}_{\ell}\right) \right]
\end{align}
where the coefficients, $a^{(n)}_{\ell}$, $b^{(n)}_{\ell}$, $c^{(n)}_{\ell}$, $d^{(n)}_{\ell}$, and $e^{(n)}_{\ell}$, come from the real-space, Doppler, gravitational redshift from the linear potential, and gravitational redshift from the halo potential, respectively. In the above, the function $\mathcal{M}$ is defined by $\mathcal{M} \equiv - 3\Omega_{\rm m0}H^{2}_{0}/(2a^{2}H)$.

To explicitly write down the the expressions of coefficients above, we introduce the dimensionless variables $x_{1} = |\bm{s}_{1}|/s$, $x_{2} = |\bm{s}_{2}|/s$, and $\cos{\theta} = \hat{\bm{s}}_{1}\cdot\hat{\bm{s}}_{2}$.
These variables can be expressed in terms of the quantities $s$, $d$, and $\mu = \hat{\bm{s}}\cdot\hat{\bm{d}}$ by
\begin{align}
x_{1} &= \frac{d}{s}\left( 1 - \frac{s}{d}\mu + \frac{1}{4}\left( \frac{s}{d}\right)^{2}\right)^{1/2} ~, \\
x_{2} &= \frac{d}{s}\left( 1 + \frac{s}{d}\mu + \frac{1}{4}\left( \frac{s}{d}\right)^{2}\right)^{1/2}~, \\
\cos{\theta} &= \frac{1-(s/d)^{2}/4}{\sqrt{\left( 1+(s/d)^{2}/4\right)^{2} - \left( s/d\right)^{2}\mu^{2}}} ~.
\end{align}

Then, the non-vanishing coefficients, $a^{(n)}_{\ell}$, $b^{(n)}_{\ell}$, $c^{(n)}_{\ell}$, $d^{(n)}_{\ell}$, and $e^{(n)}_{\ell}$ are given as follows:
\begin{align}
a^{(0)}_{2} &= -x^{2}_{1} ~, \\
b^{(0)}_{4} &= \frac{x_{1}}{7} \Biggl[ x_{2} \cos{\theta} \left( 6-14 x^{2}_{1} + 7x_{1}x_{2}\cos{\theta}\right)
-7 x_{1} (1-x^{2}_{1}) \Biggr] ~, \\
b^{(0)}_{2} &= \frac{x_{1}}{7} (6 x_{2} \cos{\theta} - 7x_{1}) ~, \\
b^{(2)}_{2} &= 6 ~, \\
c^{(1)}_{3} &= \frac{x_{1}}{5} \left(- 4 +5 x^{2}_{1} - 5 x_{1}x_{2} \cos{\theta}\right) ~, \\
c^{(1)}_{1} &= -\frac{4}{5}x_{1}~, \\
d^{(-1)}_{1} &= \frac{2}{35}\left( 7bx_{1} + 3x_{2}\cos{\theta}\right) ~,\\
d^{(-1)}_{3} &= 
\frac{1}{45}
\Biggl[
x_{2} \cos{\theta} \left(5 (9b-2) x^{2}_{1} + 12 x^{2}+5 x_{1} x_{2} \cos{\theta} \right)
\notag \\
& \qquad
+(18 b-5)x_{1}+5 (1-9 b) x^{3}_{1}
\Biggr]
~,\\
d^{(-1)}_{5} &=
\frac{1}{63}
\Biggl[
\cos{\theta} \left(6-14 x_{1}^2+7 x_{1} x_{2} \cos{\theta}\right) -7 x_{1} (1-x^{2}_{1})
\Biggr]
~,\\
d^{(2)}_{2} &= \frac{6}{x_{1}} ~,\\
e^{(-1)}_{5} &= 
\frac{1}{252}
\Biggl[ x_{1} \Biggl( 7 x_{2}^2 \Bigl(\cos (2 \theta ) \left(54 x_{1}^2-16\right)-9 x_{1} x_{2} \cos (3 \theta ) \Bigr)
\notag \\
& \qquad
-28 \left(11 x_{1}^2+4 x_{2}^2\right)+80+126 x_{1}^2 \left(2 x_{1}^2+3 x_{2}^2\right) 
\Biggr)
\notag \\
& \qquad 
- x_{2} \cos\theta  \left(-532 x_{1}^2+48 +189 x_{1}^2 \left(4 x_{1}^2+x_{2}^2\right)\right)
\Biggr]
~, \\
e^{(-1)}_{3} &= \frac{1}{45}
\Biggl[
x_{2}  \cos{\theta} \left(95 x_{1}^{2}-24\right)
\notag \\
& \qquad 
-5 x_{1} \left(-8 +11 x_{1}^{2}+4 x_{2}^{2} \cos{(2\theta)}+4 x_{2}^{2}\right)
\Biggr]
~, \\
e^{(-1)}_{1} &= \frac{4}{35} (5 x_{1}-3 x_{2} \cos{\theta}) ~.
\end{align}

\subsection{Non-vanishing multipole moments, $\xi_\ell^{\delta\gamma}$}
\label{app: multipoles}

In Sec.~\ref{sec: GI corr}, we present the analytical expression for the dipole moment of the GI correlation function. Here, we summarize all the non-vanishing multipole moments of the GI correlation function in the plane-parallel limit and leading-order wide-angle corrections as follows.

Adopting the same geometric configuration as Fig.~\ref{fig: config}, the GI correlation function depends on the separation $s=|\bm{s}|=|\bm{s}_{2}-\bm{s}_{1}|$, line-of-sight distance directing a mid point $d=|\bm{d}|=|\bm{s}_{1}+\bm{s}_{2}|/2$, and directional cosine $\mu = \hat{\bm{s}}\cdot\hat{\bm{d}}$.

We first expand the GI correlation in powers of $(s/d)$, to separate the plane parallel limit contribution from the wide-angle correction. The expression of the GI correlations is given in the sum of the contribution from the plane-parallel limit and its correction as follows:
\begin{align}
\xi^{\delta\gamma} (s,d,\mu) &=
\xi^{\delta\gamma}_{\rm pp} (s,\mu)
+ \xi^{\delta\gamma}_{\rm wa} (s,\mu) \left( \frac{s}{d} \right) + O\left( \left( \frac{s}{d} \right)^{2} \right) ~,
\end{align}
where the first and second terms on the right-hand side, respectively, correspond to the expression in the plane-parallel limit, $s/d\to 0$, and the leading-order correction to widely separated pairs.
As we have done in Eq.~(\ref{eq: decomposition}), we decompose the correlation function into four contributions:
$\xi^{\delta\gamma}_{\rm pp/wa}(s,\mu)
 = 
\xi^{(\rm r)}_{\rm pp/wa}(s,\mu)
+ \xi^{(\rm Dop)}_{\rm pp/wa}(s,\mu)
+ \xi^{(\rm grav)}_{\rm pp/wa}(s,\mu)
+ \xi^{(\rm nl)}_{\rm pp/wa}(s,\mu)$.
Further, we quantify the anisotropies of the correlation function expanded by the Legendre polynomial $\mathcal{L}_{\ell}(\mu)$:
\begin{equation}
\xi^{\delta\gamma}_{\ell}(s)
= \frac{2\ell+1}{2}\int^{1}_{-1}{\rm d}\mu\; \xi^{\delta\gamma}(s,\mu) \mathcal{L}_{\ell}(\mu)~.
\end{equation}

Then, the non-zero multipole moments from the real-space term $\xi^{({\rm r})}_{\ell}(s)$ are given by
\begin{align}
\xi^{({\rm r})}_{\rm pp,0}(s) &= -\xi^{({\rm r})}_{\rm pp,2}(s) = -\frac{2}{3}b\, b_{\rm K}\Xi^{(0)}_{2}(s) ~,
\end{align}
for the plane-parallel limit, and 
\begin{align}
\xi^{({\rm r})}_{\rm wa,1}(s) &= -\xi^{({\rm r})}_{\rm wa,3}(s) = \frac{2}{5}b\,b_{\rm K}\Xi^{(0)}_{2}(s) ~,
\end{align}
for the leading-order wide-angle correction.

The non-zero multipole moments from the Doppler term $\xi^{({\rm Dop})}_{\ell}(s)$ are given by
\begin{align}
\xi^{({\rm Dop})}_{\rm pp,0}(s) &= -\frac{2}{105}fb_{\rm K}\left( 5\Xi^{(0)}_{2}(s) - 2 \Xi^{(0)}_{4}(s)\right)
~, \label{eq: app xi_Dop_pp_0} \\
\xi^{({\rm Dop})}_{\rm pp,2}(s) &= \frac{2}{21}fb_{\rm K}\left( \Xi^{(0)}_{2}(s) + 2 \Xi^{(0)}_{4}(s)\right)
~, \\
\xi^{({\rm Dop})}_{\rm pp,4}(s) &=
- \frac{8}{35}fb_{\rm K} \Xi^{(0)}_{4}(s) ~, \label{eq: app xi_Dop_pp_4}
\end{align}
for the plane-parallel limit, and 
\begin{align}
\xi^{({\rm Dop})}_{\rm wa,1}(s) &= -\xi^{({\rm Dop})}_{\rm wa,3}(s) = \frac{2}{5}fb_{\rm K}\Xi^{(0)}_{2}(s)
~,
\end{align}
for the leading-order wide-angle correction.
The expressions of the Doppler terms in the plane-parallel limit, i.e., Eqs.~(\ref{eq: app xi_Dop_pp_0})--(\ref{eq: app xi_Dop_pp_4}), have been derived in \citet{2020MNRAS.493L.124O}.

The non-zero multipole moments from the linear gravitational redshift effect $\xi^{({\rm rel})}_{\ell}(s)$ are given by
\begin{align}
\xi^{({\rm grav})}_{\rm pp,1}(s) &= - \xi^{({\rm grav})}_{\rm pp,3}(s) = - \frac{2}{5}\mathcal{M}sb_{\rm K}\Xi^{(1)}_{3}(s) ~,
\end{align}
for the plane-parallel limit, and 
\begin{align}
\xi^{({\rm grav})}_{\rm wa,0}(s) &= - \frac{2}{15}\mathcal{M}sb_{\rm K}\left( 4 \Xi^{(1)}_{1}(s) + \Xi^{(1)}_{3}(s)\right) ~, \\
\xi^{({\rm grav})}_{\rm wa,2}(s) &= \frac{2}{105}\mathcal{M}sb_{\rm K}\left( 28 \Xi^{(1)}_{1}(s) + 13 \Xi^{(1)}_{3}(s)\right)  ~, \\
\xi^{({\rm grav})}_{\rm wa,4}(s) &=
- \frac{4}{35}\mathcal{M}sb_{\rm K} \Xi^{(1)}_{3}(s) ~,
\label{eq: app grav wa 4}
\end{align}
for the leading-order wide-angle correction.

The non-zero multipole moments from the non-linear gravitational redshift effect $\xi^{(\rm nl)}_{\ell}(s)$ are given by
\begin{align}
\xi^{(\rm nl)}_{\rm pp,1}(s) &= - \frac{\phi_{\rm halo}}{aHs}b_{\rm K}\left[ \frac{2}{5}b\,\Xi^{(-1)}_{3}(s)
+ \frac{2}{105}f\left( 7\Xi^{(-1)}_{3}(s) - 2\Xi^{(-1)}_{5}(s)\right) \right]
~, \\
\xi^{(\rm nl)}_{\rm pp,3}(s) &= \frac{\phi_{\rm halo}}{aHs}b_{\rm K} \left[ \frac{2}{5}b\,\Xi^{(-1)}_{3}(s)
 + \frac{2}{45}f\left( 3\Xi^{(-1)}_{3}(s) + 2\Xi^{(-1)}_{5}(s)\right) \right]
~, \\
\xi^{(\rm nl)}_{\rm pp,5}(s) &= 
- \frac{8}{63}f\frac{\phi_{\rm halo}}{aHs}b_{\rm K} \Xi^{(-1)}_{5}(s) ~,
\label{eq: app nl pp 5}
\end{align}
for the plane-parallel limit, and 
\begin{align}
\xi^{(\rm nl)}_{\rm wa,0}(s) &= - \frac{\phi_{\rm halo}}{aHs}b_{\rm K}\Biggl[ \frac{2}{945} \Bigl( 18(3+7b)\Xi^{(-1)}_{1}(s)
\notag \\
& \qquad
- 7(-8+9b)\Xi^{(-1)}_{3}(s) + 2\Xi^{(-1)}_{5}(s)\Bigr)
\notag \\
& \qquad 
+ \frac{2}{945}f\left( 72 \Xi^{(-1)}_{1}(s) - 7\Xi^{(-1)}_{3}(s) +2\Xi^{(-1)}_{5}(s)\right) \Biggr]
~, \\
\xi^{(\rm nl)}_{\rm wa,2}(s) &= 
-\frac{\phi_{\rm halo}}{aHs}b_{\rm K}
\Biggl[
- \frac{2}{945} \Bigl( 18(3+7b)\Xi^{(-1)}_{1}(s)
\notag \\
& \qquad
+ (44-9b)\Xi^{(-1)}_{3}(s) - 10\Xi^{(-1)}_{5}(s)\Bigr)
\notag \\
& \qquad 
- \frac{2}{945}f\left( 72 \Xi^{(-1)}_{1}(s) + 107\Xi^{(-1)}_{3}(s) -10\Xi^{(-1)}_{5}(s)\right) 
\Biggr]
 ~, \\
\xi^{(\rm nl)}_{\rm wa,4}(s) &= - \frac{\phi_{\rm halo}}{aHs}b_{\rm K} \Biggl[ \frac{4}{315} \left( (-2+9b)\Xi^{(-1)}_{3}(s) - 2\Xi^{(-1)}_{5}(s)\right) 
\notag \\
& \qquad
+ \frac{4}{3465}f\left( 209\Xi^{(-1)}_{3}(s) -52\Xi^{(-1)}_{5}(s)\right) 
\Biggr]
~, \\
\xi^{(\rm nl)}_{\rm wa,6}(s) &=
- \frac{8}{231}f\frac{\phi_{\rm halo}}{aHs}b_{\rm K} \Xi^{(-1)}_{5}(s) ~, \label{eq: app nl wa 6}
\end{align}
for the leading-order wide-angle correction.
In the above, we omitted the time dependence from the argument of each function. The function $\Xi^{(n)}_{\ell}(s)$ is defined as
\begin{equation}
\Xi^{(n)}_{\ell}(s) = \int\frac{k^{2}{\rm d}k}{2\pi^{2}}\frac{j_{\ell}(ks)}{(ks)^{n}} P_{\rm L}(k) ~,
\end{equation}
with the function $P_{\rm L}$ being the linear power spectrum at a given redshift.

It is worth noting that the multipole moments presented above satisfy the following relations:
\begin{equation}
\sum_{\ell}\xi^{({\rm x})}_{{\rm pp/wa}, \ell}(s) = 0,\quad {\rm x} = \{{\rm r,\, Dop,\, grav,\, nl}\}.
\label{eq:moment_sum}
\end{equation}
These relations come from the fact that the correlation function of each contribution involves the factor of $(1-\mu^2)$ arising from the projected ellipticity field (Eqs.~(\ref{eq:xi_r_pp})--(\ref{eq: xi eps wa})), and can be simply derived as follows. For a correlation function $\xi(s,\mu)$ involving the factor $(1-\mu^2)$, we generally have $\xi(s,\pm1)=0$. 
This implies that when expanding the function $\xi$ 
as $\xi(s,\mu)=\sum_\ell \xi_\ell(s)\,\mathcal{L}_\ell(\mu)$, one obtains 
$\sum_\ell \xi_\ell(s)=0$ and 
$\sum_\ell (-1)^\ell\xi_\ell(s)=0$, where we used the fact that 
$\mathcal{L}_\ell(1)=1$ and $\mathcal{L}_\ell(-1)=(-1)^\ell$. These relations are equivalently rewritten with $\sum_{\ell={\rm even}} \xi_\ell(s)=0$ and $\sum_{\ell={\rm odd}} \xi_\ell(s)=0$. 
Now recalling that multipole moments of
the wide-angle correction and plane-parallel limit appear non-vanishing for either even or odd $\ell$ in each contribution, we arrive at Eq.~(\ref{eq:moment_sum}).

\subsection{Non-vanishing spherical harmonic coefficients, $P_{\ell,m}^{\rm XY}$}
\label{app: coefficients Ylm}

As we have seen in Sec.~\ref{sec: covariance}, the GI correlation in the plane-parallel limit generally depends not only on the directional cosine $\mu$ but also the azimuthal angle $\varphi$ (see Eq.~(\ref{eq: def coordinate})). Their explicit dependencies can be conveniently characterized by expanding the Fourier counterpart of the correlation function with the spherical harmonics $Y_{\ell,m}(\hat{\bm{k}})$ as follows:
\begin{align}
\xi^{\rm XY}(\bm{s}) = \int\frac{{\rm d}^3k}{(2\pi)^{3}}e^{-i\bm{k\cdot\bm{s}}} \sum_{\ell, m}P^{\rm XY}_{\ell,m}(k) Y_{\ell,m}(\hat{\bm{k}}) ~,
\end{align}
with $\bm{s} = \bm{s}_{2} - \bm{s}_{1}$.

Here, we summarize the non-zero spherical harmonic coefficients used in the expression of the covariance matrix.
The non-zero coefficients of the GI correlation function are given by
\begin{align}
P^{\delta\gamma ({\rm r})}_{2,\pm 2}(k) &= -i^{2} 2b_{\rm K}b \sqrt{\frac{2\pi}{15}} P_{\rm L}(k) ~, \\
P^{\delta\gamma ({\rm Dop})}_{2,\pm 2}(k) &= -i^{2} \frac{2}{7}b_{\rm K} f \sqrt{\frac{2\pi}{15}} P_{\rm L}(k) ~, \\
P^{\delta\gamma ({\rm Dop})}_{4,\pm 2}(k) &= i^{4} \frac{4}{21}b_{\rm K} f\sqrt{\frac{2\pi}{5}} P_{\rm L}(k) ~,\\
P^{\delta\gamma ({\rm grav})}_{3,\pm 2}(k) &= -i^{3}2b_{\rm K}s\mathcal{M} \sqrt{\frac{2\pi}{105}} (ks)^{-1} P_{\rm L}(k)~, \\
P^{\delta\gamma (\rm nl)}_{3,\pm 2}(k) &= - i^{3}\frac{2}{3}b_{\rm K}\frac{\phi_{\rm halo}}{aHs}(3b+f) \sqrt{\frac{2\pi}{105}} (ks) P_{\rm L}(k)~,\\
P^{\delta\gamma (\rm nl)}_{5,\pm 2}(k) &= i^{5}\frac{4}{3}b_{\rm K}\frac{\phi_{\rm halo}}{aHs} f \sqrt{\frac{2\pi}{1155}} (ks)P_{\rm L}(k) ~.
\end{align}
The non-zero coefficients of the GG correlation function are given by
\begin{align}
P^{\delta\delta}_{0,0}(k) & = i^{0}\sqrt{4\pi}\left( b^{2} + \frac{2}{3}b\, f + \frac{1}{5}f^{2} \right)P_{\rm L}(k) ~, \\
P^{\delta\delta}_{2,0}(k) & = -i^{2}\sqrt{\frac{4\pi}{5}}\left( \frac{4}{3}b\, f + \frac{4}{7}f^{2} \right) P_{\rm L}(k)~, \\
P^{\delta\delta}_{4,0}(k) & = i^{4}\sqrt{\frac{4\pi}{9}}\frac{8}{35}f^{2} P_{\rm L}(k)~.
\end{align}
The non-zero coefficients of the II correlation function are given by
\begin{align}
P^{\gamma\gamma}_{0, 0}(k) &= i^{0} b^{2}_{\rm K}\frac{8\sqrt{\pi}}{15} P_{\rm L}(k) ~, \\
P^{\gamma\gamma}_{2, 0}(k) &= i^{2}b^{2}_{\rm K}\frac{16}{21} \sqrt{\frac{\pi}{5}} P_{\rm L}(k)~, \\
P^{\gamma\gamma}_{4, 0}(k) &= i^{4} b^{2}_{\rm K}\frac{8\sqrt{\pi}}{105} P_{\rm L}(k) ~, \\
P^{\gamma\gamma}_{4, \pm 4}(k) &= i^{4}b^{2}_{\rm K} \frac{4}{3} \sqrt{\frac{2\pi}{35}} P_{\rm L}(k)~.
\end{align}

\section{Covariance matrix for GI correlation}
\label{app: covariance}

To give an analytical formula for the Gaussian covariance, let us define the estimator for the dipole moment of the cross-correlation function.
Assuming the plane-parallel limit, the cross-correlation function can be written as a function of the separation vector between two objects, $\bm{s} = \bm{s}_{2} - \bm{s}_{1}$:
\begin{align}
\hat{\xi}^{\rm XY}(\bm{s}) = \int\frac{{\rm d}^{3}r}{V}\, {\rm X}\left(\bm{r}-\frac{\bm{s}}{2}\right){\rm Y}\left(\bm{r}+\frac{\bm{s}}{2}\right) ~, \label{eq: estimator}
\end{align}
where the quantity $V$ stands for the observed volume.
Here, the functions ${\rm X}$ and ${\rm Y}$ stand for the observed galaxy density fluctuation ($\delta$) or galaxy's ellipticity function ($\gamma_{+/\times}$), which satisfy
\begin{align}
\Braket{ {\rm X}(\bm{r}) {\rm Y}(\bm{r}')} = \xi^{\rm XY}(\bm{r}'-\bm{r}) + N^{\rm X} \delta^{\rm K}_{\rm X,Y} \delta^{\rm D}(\bm{r}'-\bm{r}) ~.
\end{align}
The second term represents the noise contribution, and the quantity $N^{\rm X}$ indicates $N^{\delta} = 1/n_{\rm g}$ for the galaxy density field ($X=\delta$) or $N^{\gamma} = \sigma^{2}_{\rm shape}/n_{\rm g}$ for the ellipticity field ($X=\gamma$), where $n_{\rm g}$ and $\sigma_{\rm shape}$ are the galaxy number density and root mean square of the galaxy's ellipticity, respectively. Note that the noise term becomes non-zero only in the self-correlation case, i.e., ${\rm X}={\rm Y}$ and $\bm{r} = \bm{r}'$.

Let us define the covariance of the correlation function by
\begin{align}
{\rm COV}(\bm{s},\bm{s}')
= \Braket{\hat{\xi}^{\rm XY}(\bm{s})\hat{\xi}^{\rm XY}(\bm{s}')} - \Braket{\hat{\xi}^{\rm XY}(\bm{s})}\Braket{\hat{\xi}^{\rm XY}(\bm{s}')}
~, \label{eq: def covariance}
\end{align}
and then, substituting Eq.~(\ref{eq: estimator}) into Eq.~(\ref{eq: def covariance}), we have
\begin{align}
& {\rm COV}(\bm{s},\bm{s}') \notag \\
& = \int\frac{{\rm d}^{3}r''}{V}
\Biggl[
\xi^{\rm XX}\left( \frac{1}{2}\bm{s} - \frac{1}{2}\bm{s}' + \bm{r}'' \right)
\xi^{\rm YY}\left( \frac{1}{2}\bm{s}' - \frac{1}{2}\bm{s} + \bm{r}'' \right)
\notag \\
& \qquad
+ 
\xi^{\rm XY}\left( \frac{1}{2}\bm{s} + \frac{1}{2}\bm{s}' + \bm{r}'' \right)
\xi^{\rm XY}\left( \frac{1}{2}\bm{s}' + \frac{1}{2}\bm{s} - \bm{r}'' \right)
\Biggr]
\notag \\
& \qquad +
\frac{1}{V}
\Biggl[
\xi^{\rm X}\left( \bm{s} - \bm{s}'\right) N^{\rm YY}
+
\xi^{\rm Y}\left( \bm{s}' - \bm{s} \right) N^{\rm XX}
\Biggr]
\notag \\
& \qquad +
\frac{1}{V}
\Biggl[
N^{\rm X}N^{\rm Y}\delta_{\rm D}(\bm{s}-\bm{s}')
\Biggr]
~. \label{eq: cov def}
\end{align}
To further compute Eq~(\ref{eq: cov def}), we use the expression of the power spectrum:
\begin{align}
\xi^{\rm XY}(\bm{s}) & = \int\frac{{\rm d}^3k}{(2\pi)^{3}}e^{-i\bm{k\cdot\bm{s}}}P^{\rm XY}(\bm{k}) ~, \label{eq: xi_XY to P_XY}
\end{align}
with $\bm{s} = \bm{s}_{2} - \bm{s}_{1}$. The power spectrum satisfies the relation: $P^{\rm XY}(\bm{k}) = P^{\rm YX}(-\bm{k})$.

Unlike the GG cross-correlation, the GI correlation function generally depends not only on the directional cosine $\mu$ but also on the azimuthal angle $\phi$.
We then expand the power spectrum by using the spherical harmonics $Y_{\ell,m}(\hat{\bm{k}})$:
\begin{align}
P^{\rm XY}(\bm{k}) = \sum_{\ell, m}P^{\rm XY}_{\ell,m}(k)\, Y_{\ell,m}(\hat{\bm{k}}) ~. \label{eq: def P_ellm}
\end{align}
We summarize the coefficients $P^{\rm XY}_{\ell,m}$ in Appendix~\ref{app: coefficients Ylm} in the plane-parallel limit.

Then, substituting Eqs.~(\ref{eq: xi_XY to P_XY}) and (\ref{eq: def P_ellm}) into Eq.~(\ref{eq: cov def}), a straightforward but lengthy calculation lead to
\begin{align}
& {\rm COV}(\bm{s},\bm{s}')
\notag \\
& = 
\frac{1}{V}
\int\frac{k^{2}{\rm d}k}{2\pi^{2}}
\sum_{L,L'}
j_{L}(ks)j_{L'}(ks')
\sum_{\ell_{1},m_{1}}
\sum_{\ell_{2},m_{2}}
\sum_{M,M'}
\mathcal{G}^{(\ell_{1},m_{1})(\ell_{2},m_{2})}_{L,M,L',M'}
\notag \\
& \qquad \times 
\left[ P^{\rm XX}_{\ell_{1},m_{1}}(k)P^{\rm YY}_{\ell_{2},m_{2}}(k)
+ (-1)^{L'}P^{\rm XY}_{\ell_{1},m_{1}}(k)P^{\rm XY}_{\ell_{2},m_{2}}(k)
\right]
\notag \\
& \qquad \times 
\sqrt{\frac{4\pi}{2L+1}}\sqrt{\frac{4\pi}{2L'+1}}
Y_{L,M}^{*}(\hat{\bm{s}})Y_{L',M'}^{*}(\hat{\bm{s}}')
\notag \\
&
+ \frac{1}{V}\int\frac{k^{2}{\rm d}k}{2\pi^{2}}
\sum_{L,L'}j_{L}(ks)j_{L'}(ks')
\sum_{\ell_{3},m_{3}}\sum_{M,M'}
\mathcal{G}^{(\ell_{3},m_{3})}_{L,M,L',M'}
\notag \\
& \qquad \times
\left[ P^{\rm XX}_{\ell_{3},m_{3}}(k) N^{\rm Y} + P^{\rm YY}_{\ell_{3},m_{3}}(k) N^{\rm X} \right]
\notag \\
& \qquad \times
\sqrt{\frac{4\pi}{2L+1}}\sqrt{\frac{4\pi}{2L'+1}}Y^{*}_{L,M}(\hat{\bm{s}})Y^{*}_{L',M'}(\hat{\bm{s}}')
\notag \\
&
+ \frac{N^{\rm X}N^{\rm Y} }{s^{2} L_{\rm p} V}\delta^{\rm K}_{s,s'}\frac{\delta^{\rm D}(\theta - \theta')}{\sin{\theta}}\delta^{\rm D}(\varphi - \varphi')~, \label{eq: cov general}
\end{align}
Here the coefficients $\mathcal{G}^{(\ell_{1},m_{1})(\ell_{2},m_{2})}_{L,M,L',M'}$ and $\mathcal{G}_{L,M,L',M'}^{(\ell_{3},m_{3})}$ are defined and are expressed in terms of the Wigner 3-$j$ symbols below:
\begin{align}
\mathcal{G}^{(\ell_{1},m_{1})(\ell_{2},m_{2})}_{L,M,L',M'} & \equiv 
i^{L'-L}
\sqrt{\frac{2\ell_{1}+1}{4\pi}}\sqrt{\frac{2\ell_{2}+1}{4\pi}}
\notag \\
&\quad  \times 
(2L+1)(2L'+1)
\sum_{\ell_{3},m_{3}}
(-1)^{m_{3}} (2\ell_{3}+1)
\notag \\
&\quad  \times 
\left(
\begin{array}{ccc}
\ell_{3} & L & L' \\
0 & 0 & 0
\end{array}
\right)
\left(
\begin{array}{ccc}
\ell_{3} & \ell_{1} & \ell_{2} \\
0 & 0 & 0
\end{array}
\right)
\notag \\
& \quad \times 
\left(
\begin{array}{ccc}
\ell_{3} & L & L' \\
m_{3} & M & M'
\end{array}
\right)
\left(
\begin{array}{ccc}
\ell_{3} & \ell_{1} & \ell_{2} \\
-m_{3} & m_{1} & m_{2}
\end{array}
\right) ~, \label{eq: def calG l1l2}
\\
\mathcal{G}_{L,M,L',M'}^{(\ell_{3},m_{3})} & \equiv 
i^{L'-L} (2L+1)(2L'+1)\sqrt{\frac{2\ell_{3}+1}{4\pi}}
\notag \\
& \quad \times
\left(
\begin{array}{ccc}
L & L' & \ell_{3} \\
0 & 0 & 0
\end{array}
\right)
\left(
\begin{array}{ccc}
L & L' & \ell_{3} \\
M & M' & m_{3}
\end{array}
\right) ~. \label{eq: def calG l3}
\end{align}

When observationally characterizing anisotropies of the correlation functions in a finite volume survey, one often uses the Legendre polynomials and applies the multipole expansion, fixing the azimuthal angle. Thus, to obtain the covariance of the multipole moment, especially dipole moment, we take an average over the directional cosine, $\mu$, weighting with the Legendre polynomials, $\mathcal{L}_{\ell}(\mu)$:
\begin{align}
{\rm COV}_{\ell}(\bm{s},\bm{s}') & =
\left( \frac{2\ell+1}{2} \right)^{2}
\int^{1}_{-1}{\rm d}\cos{\theta}
\int^{1}_{-1}{\rm d}\cos{\theta'}
\notag \\
& \qquad \times 
\mathcal{L}_{\ell}(\cos{\theta})
\mathcal{L}_{\ell}(\cos{\theta'})
\, 
{\rm COV}(\bm{s},\bm{s}') ~.
\end{align}
Then, we obtain the covariance of the multipole moment, ${\rm COV}_{\ell}(\bm{s},\bm{s}')$, by
\begin{align}
& {\rm COV}_{\ell}(\bm{s},\bm{s}')
\notag \\
&
= 
\frac{1}{V}
\int\frac{k^{2}{\rm d}k}{2\pi^{2}}
\sum_{L,L'}
j_{L}(ks)j_{L'}(ks')
\notag \\
& \quad
\times \sum_{\ell_{1},m_{1}}
\sum_{\ell_{2},m_{2}}
\sum_{M,M'}
\Bigl[ P^{\rm XX}_{\ell_{1},m_{1}}(k)P^{\rm YY}_{\ell_{2},m_{2}}(k)
+ (-1)^{L'}P^{\rm XY}_{\ell_{1},m_{1}}(k)P^{\rm XY}_{\ell_{2},m_{2}}(k)\Bigr]
 \notag \\
& \quad
\times
\mathcal{G}_{L,M,L',M'}^{(\ell_{1},m_{1})(\ell_{2},m_{2})}
\mathcal{I}_{\ell,L,M}
\mathcal{I}_{\ell,L',M'}
e^{-iM\phi -iM'\phi'}
\notag \\
& + \frac{1}{V}\int\frac{k^{2}{\rm d}k}{2\pi^{2}}
\sum_{L,L'}j_{L}(ks)j_{L'}(ks')
\notag \\
& \quad 
\times \sum_{\ell_{3},m_{3}}\sum_{M,M'}
\left[ P^{\rm XX}_{\ell_{3},m_{3}}(k) N^{\rm Y} + P^{\rm YY}_{\ell_{3},m_{3}}(k) N^{\rm X} \right]
\notag \\
& \quad 
\times 
\mathcal{G}_{L,M,L',M'}^{(\ell_{3},m_{3})}
\mathcal{I}_{\ell,L,M}
\mathcal{I}_{\ell,L',M'}
e^{-iM\varphi -iM'\varphi'}
\notag \\
& + \frac{N^{\rm X}N^{\rm Y} }{s^{2} L_{\rm p} V} \frac{2\ell+1}{2} \delta^{\rm K}_{s,s'} \delta^{\rm D}(\varphi -\varphi')~.
\label{eq: cov ell}
\end{align}
In the above, we define the (real number) quantity $\mathcal{I}_{\ell,L,M}$ \citep[see e.g.,][for the relevant integrals]{Salem_1989}:
\begin{align}
\mathcal{I}_{\ell,L,M} = 
\sqrt{\frac{4\pi}{2L+1}}
\frac{2\ell+1}{2}
e^{iM\varphi}\int^{1}_{-1}{\rm d}\cos{\theta}\,
Y_{L,M}^{*}(\hat{\bm{s}})
\mathcal{L}_{\ell}(\cos{\theta}) ~. \label{eq: def calI}
\end{align}
This function, $\mathcal{I}_{\ell,L,M}$, arises from the directional cosine integral of the product of the spherical harmonics and Legendre polynomial, and has the following properties:
\begin{align}
\mathcal{I}_{\ell,L,M} &= 0 \quad \mbox{$\ell > L$ and $M$ is even or $\ell + L- M$ is odd} ~,\\
\mathcal{I}_{\ell,L,0} &= \delta^{\rm K}_{\ell,L} ~,\\
\mathcal{I}_{\ell,L,-M} &= (-1)^{M}\mathcal{I}_{\ell,L,M} ~.
\end{align}
Eq.~(\ref{eq: cov ell}) provides us with a covariance matrix of the GI correlation function, taking properly its angular dependence into account.

\bsp
\label{lastpage}
\end{document}